\documentclass[onecolumn]{elsart}
\usepackage[square,comma,sort&compress,numbers]{natbib}
\usepackage{graphicx}
\usepackage{amsmath}
\usepackage{amssymb}
\usepackage{amsfonts}
\usepackage{hyperref}
\usepackage{bm}
\usepackage[usenames,dvips]{color}
\journal{the Journal of Computational Physics}
\begin{document}
\begin{frontmatter}

\title{Multidomain Spectral Method for
       the Helically Reduced Wave Equation}

\author{Stephen~R.~Lau}
\ead{lau@dam.brown.edu}
\address{
Department of Mathematics and Statistics, 
University of New Mexico, 
Albuquerque, NM 87111}
\address{and}
\address{Division of Applied Mathematics,
Brown University,
Providence, RI 02912}
\author{Richard H.~Price}
\ead{rprice@phys.utb.edu}
\address{
Center for Gravitational Wave Astronomy, 
Department of Physics and Astronomy,
University of Texas at Brownsville, 
Brownsville, TX 78520}

\begin{abstract}
We consider the 2+1 and 3+1 scalar wave equations reduced via a helical 
Killing field, respectively referred to as the 2--dimensional and 
3--dimensional helically reduced wave equation (HRWE). The HRWE serves 
as the fundamental model for the mixed--type PDE arising in the periodic 
standing wave (PSW) approximation to binary inspiral. We present a method 
for solving the equation based on domain decomposition and spectral 
approximation. Beyond describing such a numerical method for solving 
strictly linear HRWE, we also present results for a nonlinear scalar 
model of binary inspiral. The PSW approximation has already been 
theoretically and numerically studied in the context of the 
post--Minkowskian gravitational field, with numerical simulations 
carried out via the ``eigenspectral method.'' Despite its name, the 
eigenspectral technique does feature a finite--difference component, 
and is lower--order accurate. We intend to apply the numerical method 
described here to the theoretically well--developed post--Minkowski PSW
formalism with the twin goals of spectral accuracy and the coordinate
flexibility afforded by global spectral interpolation.  
\end{abstract}
\end{frontmatter}
\section{Introduction and preliminaries}

\subsection{Periodic standing wave (PSW) approximation}

The development of gravitational wave detectors has spurred interest
in the binary inspiral of compact astrophysical objects, in particular
black holes.  The challenge of computationally solving Einstein's
equations for such a system has become the focus of many groups
throughout the world both for its importance in gravitational wave
astronomy and for its role in advancing the understanding of highly
dynamical strongly curved spacetimes. Very 
recently\cite{Pretorius,UTB,UTBhangup,UTBspinflip,NASA,CalCorn,Jena} 
a number of techniques have been found to stabilize codes, so that
computational evolution of Einstein's equations can track the last few
orbits of binary inspiral.

There is, of course, more to the problem than the last few orbits.
For orbiting objects with mass $M$ and separation $a$, the
characteristic measure of the nonlinearity of their gravitational
interaction is $GM/ac^2$ where $G,c$ are the gravitational constant
and the speed of light. When this measure is small, the dynamics and
the generation of gravitational waves can be found using
post-Newtonian methods, an analytic approach in which Einstein's
theory is, in effect, expanded treating $GM/ac^2$ as a perturbation
parameter.

With methods available for the small-$GM/ac^2$ early stage, and the
large-$GM/ac^2$ last few orbits, what remains needed is an effective method
for treating the intermediate epoch, the stage of inspiral in which
$GM/ac^2$ is too large for post-Newtonian methods, but in which many
orbits remain. When more than a few orbits still remain, an
accurate numerical evolution of Einstein's equations will be too 
computationally expensive, at least in the near future.

The periodic standing wave (PSW) approach is an approximation scheme
for dealing effectively with this intermediate inspiral epoch. The
scheme is based on the fact that during this epoch the orbiting
inspiral is very much orbit, and only slightly inspiral. That is, the
change in the orbital radius is small for each orbit. (This is, in
fact, a criterion for many orbits to remain, and for accurate
computational evolution to be daunting.) In the PSW 
approach the motion of the
sources and the fields are assumed to be helically symmetric, that is,
all quantities are rigidly rotating in the sense that a change in time
by $\Delta t$ is equivalent to a change in azimuthal angle $\varphi$
according to $\Delta\varphi\rightarrow-\Omega\Delta t$, where $\Omega$
is a constant, the angular velocity with which the fields rigidly
rotate.

The imposition of this helical symmetry vastly changes the nature of
the mathematical and computational problem. Prior to helical reduction, 
the problem is that of evolving forward in time a hyperbolic problem (more
precisely, a problem that can be cast in hyperbolic form). The
imposition of helical symmetry reduces by one the number of
independent variables and, more important, changes the problem from a
hyperbolic problem to a mixed one, a problem with a region of the manifold
(near the rotation axis) in which the equations are elliptic, and an
outer region in which the equations are hyperbolic. The boundary
conditions for this problem are also unusual.  One can have the
presence of a source represented by inner Dirichlet boundary
conditions on two small topological spheres just outside the location
of the sources, the compact astrophysical objects. Alternatively, one
can include the objects as explicit inhomogeneities in the
equations. The other boundary conditions on the problem represent the
radiation in the distant wave zone, and require some discussion.

An important feature of this problem in general relativity is the
conservation of the total energy of the system.  If energy is leaving
in the form of outgoing gravitational radiation, then the orbital
motion cannot be helically symmetric; the radius must decrease. For
fields other than gravity one could invoke a force, some {\it deux ex
machina}, to keep the orbits unchanging, and have that force not
couple to the field being studied.  This certainly can be, and has
been done in model
problems\cite{WheKrivPri,WBLP,PriceCQG,Andradeetal,BOP2005} but, in
principle, cannot be done in general relativity. In Einstein's
gravitation all forces couple to gravity. This is dealt with by
computing a standing wave solution of the helically symmetric problem,
that is, by imposing standing wave outer boundary conditions. From the
standing wave exact solution, an approximate solution is then
extracted to the physical problem of outgoing waves.

By ``standing waves'' in a {\em linear} theory, we mean here the
average of the outgoing wave solution and the ingoing wave
solution. No clear meaning exists for standing waves in a nonlinear theory.  
To define our standing waves, we choose an iterative solution technique
for the nonlinear problem in which an average of ingoing and outgoing
solutions is taken at each step of iteration.\footnote{Another
extension of ``standing wave'' to nonlinear theories has been
discussed in the literature\cite{WBLP}, using the minimum amplitude of
the multipole in every multipole mode.}
We emphasize here that in each iteration of the PSW
problem it is an outgoing (or, trivially different, ingoing) problem
that is solved.  The present paper will therefore focus on the details of the
efficient computation of a problem with outgoing radiation boundary
conditions.

Hyperbolic problems with Dirichlet boundary conditions are known not
to be well posed.  We take a pragmatic approach to whether our
mixed-type problems with radiative conditions are well posed. For one
thing, the problem arises in what would appear to be a physically
well-specified problem; heuristically the problems of nonuniqueness
for the Dirichlet problem are removed by the radiation boundary
conditions.  For another thing, no fundamental instability has been
encountered in seeking numerical solutions of our problems. Further
indirect evidence that the problem is well posed can be found in the
work of Torre, who has shown that a closely related mixed-type linear
problem is well-posed if the boundary conditions are of an admissible
type that includes Sommerfeld outgoing conditions\cite{Torre1,Torre2}.

The strategy of the PSW approach is to solve the helically symmetric
binary problem computationally, but otherwise without approximation.
From that ``exact'' helically symmetric solution, an approximation is
extracted for the physical problem.  The extraction of an outgoing
approximation is tantamount to treating the nonlinear standing wave
solution as if it were an average of the ingoing and outgoing
solutions. (For details of extraction, see\cite{BOP2005}.) The reason
that this is an excellent approximation (as is confirmed by
computations with model problems) is that the regions of the physical
manifold in which nonlinearities are strong are those near the
sources, and in these regions the solution is extremely insensitive to
the boundary conditions (ingoing, outgoing, standing wave) in the
distant weak field wave region. Since the  ingoing and outgoing solutions 
are nearly identical in this region they can be averaged (or simply replaced
by either the ingoing or the outgoing solution).
Where the ingoing/outgoing solutions
are very different, the theory is approximately linear and hence again
the ingoing and outgoing fields can be averaged.
This feature, the separation of the region of nonlinearity and the
region of waves, is closely related to the argument, given below, that
a multidomain spectral method should have advantages for PSW type
problems even beyond the advantages it has demonstrated for purely
elliptic initial-value problems in the work of Pfeiffer 
{\it et al.}\cite{Pfeifferthesis,PKSTelliptic}, or its use in 
evolution codes by the Caltech-Cornell collaboration\cite{Boyleetal}.

Successful solutions of the PSW approximation will serve a variety of
purposes.  As discussed above, it will provide the bridge between the
post-Newtonian methods and numerical evolution for binary orbits; it
will provide near optimal starting points for the numerical evolution
of the last few orbits; it may give a useful testbed for studying
radiation reaction; solution of the PSW standing wave problem will
give a new class of solutions to Einstein's theory.

Work on PSW computations has already taken several steps using the
``eigenspectral method,'' an approximate numerical method based on
coordinates well adapted to the geometry both close to the sources and
in the radiation region. This method, in effect, keeps only the
features of the solution that are most important to the structure of
the near-source fields and to the radiation, and has been used so far
to study nonlinear toy models\cite{WheKrivPri,WBLP,BOP2005},
linearized general relativity\cite{BBP2006}, and the post-Minkowski
extension of linearized gravity\cite{BBHP2006}. This method is
remarkable for its simplicity in giving approximate solutions, but is
ill suited to the high accuracy needed for several purposes. In
addition the eigenspectral technique uses a finite-difference
component, and provides numerical answers on a specific grid of
points in the space of the physical problem. Its results, therefore,
require a difficult interpolation from a nonuniform grid if they 
are to be compared with other results or used as initial conditions 
for evolution codes. The multidomain spectral method we will describe 
has neither shortcoming.

To minimize the complexity of issues not directly relevant to the
multidomain spectral method, we choose as a specific target problem,
more-or-less the  problem of
Refs.~\cite{WheKrivPri,WBLP,BOP2005}, that of a nonlinear scalar
field.  The physical problem, of course, is one with three spatial
coordinates, but for simplicity both of exposition and of computation,
we work here with the two-dimensional helically reduced wave equation,
hereafter 2d HRWE. The Cauchy problem for this physical model would
involve three independent variables (two spatial, one time), but the 
helical reduction means we are solving on a manifold of two 
dimensions. We will also, in Sec.~\ref{numerical}, present some
preliminary results for the 3d HRWE, and in the concluding section
will discuss application of the method to the 3d HRWE problem.
The physical picture is that of 
two compact scalar charges moving on
circular orbits of radius $a$, with angular velocity $\Omega$, and emitting
helically symmetric outgoing waves. The wave speed $c$ will be taken to
be unity. The scalar field satisfies a 2d HRWE of the form  
\begin{equation}
L\psi + \eta h(\psi) = g\,,
\label{introHRWE}
\end{equation}
on the region outside the two sources and inside a large radius circle
(spherical surface for 3d HRWE) on which outgoing conditions are
imposed. Here $L$ is a mixed-type linear HRWE operator (the helically reduced
d'Alembertian), $h$ is a nonlinear function of $\psi$, and the $\eta$
parameter controls the strength of the nonlinearity. The inhomogeneity
$g$ can be used to represent explicit sources. In our 2d HRWE
development below we will use inner boundary conditions, not explicit
sources, so $g$ will be zero; a nonzero $g$ will be used only in
the consideration of the 3d HRWE in the concluding section.

We make one more simplifying choice. We take the solution to have the
symmetry $\psi(-x,-y) = -\psi(x,y)$, a choice equivalent to taking the
sources to have opposite scalar charge.  By setting the total charge
equal to zero, we eliminate the bothersome logarithmic monopole that
would diverge at large distances.  Of course, for gravitation theory
there is only charge of a single sign and the monopole cannot be set
to zero, but the real physical problem is that for three spatial
dimensions, in which the monopole falls off with distance from the
sources, and is an acceptable complication.

\subsection{Multidomain spectral method} 
\begin{figure}
\centering
\scalebox{0.6}{\includegraphics{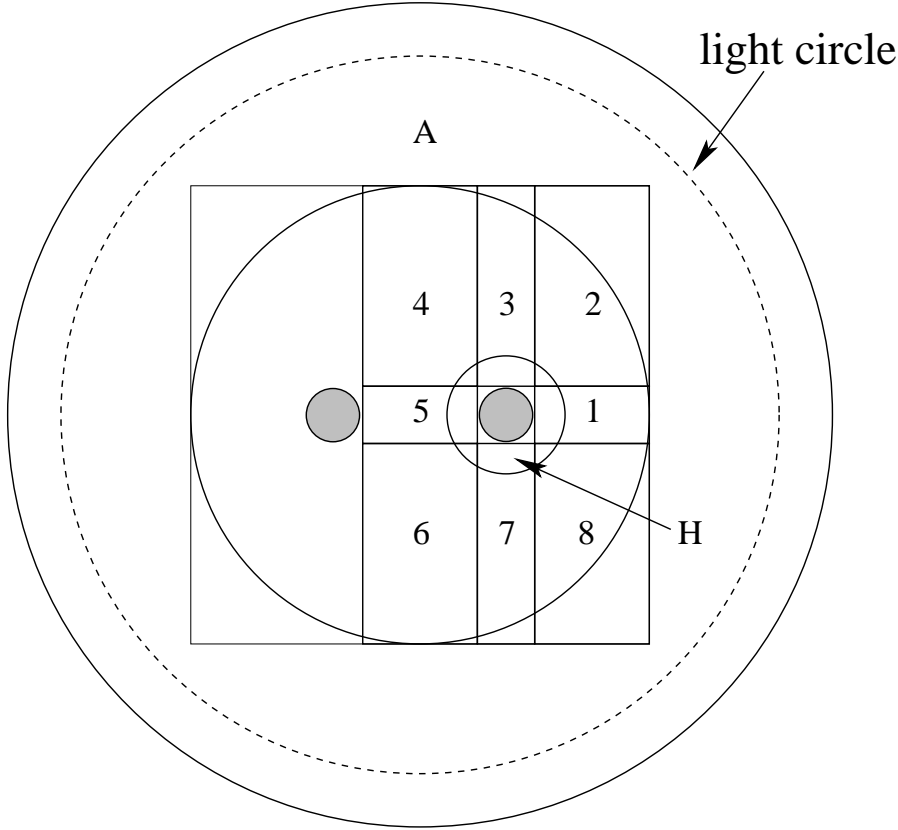}}
\caption{{\sc The 2d HRWE problem and its domain decomposition.} 
The circles that are the outer boundaries of the darkened source regions carry 
Dirichlet boundary conditions representing the imprint of the sources. Region
$H$, between the two small circles concentric with the source, is the
inner annular domain.  Domain $A$ between two large solid circles is
the outer annular domain.  
The dashed circle in domain $A$ is 
the light circle.
 }
\label{domaindecomp}
\end{figure}

The ``two center'' domain of the 2d HRWE problem is illustrated in
Fig.~\ref{domaindecomp}. This figure shows how two annular domains,
$H$ and $A$ along with eight rectangular domains, fully cover the
physical problem.  Note that the symmetry $\psi(-x,-y) = -\psi(x,y)$
means that annulus $H$ represents the information around both the
source on the right, and that on the left.  Similarly, rectangular
domains 1,2,3,7,8 carry the information about the solution in symmetry
related regions on the left side of the physical problem. Using 
the aforementioned symmetry, we could do away with region 6, which 
is equivalent to region 4. Nevertheless, we have chosen to keep 
region 6 so that our code can be tested on elliptic problems which, 
when posed on the inner region spanned by the rectangles, need not
be of definite parity. 
The annular domain $A$ extends between the two solid circles in
Fig.~\ref{domaindecomp}. At its outer boundary we impose
Sommerfeld-like outgoing radiation boundary conditions to be described
below.  For most radiation conditions to be applicable, the outer boundary
of $A$ must be at least several wavelengths away from the
sources. For a typical choice of $\Omega$ this means that
the radius of the outer boundary must be 10 times or more larger than
the separation between the source regions.  The dashed circle in
Fig.~\ref{domaindecomp} is the ``light circle,'' the boundary between
the inner region in which the operator $L$ is elliptic, and the outer
region in which $L$ is hyperbolic. Typically the radius of the the
light circle will be at least several times larger than the
separation of the centers of the source domains.
\begin{figure}
\centering
\scalebox{0.55}{\includegraphics{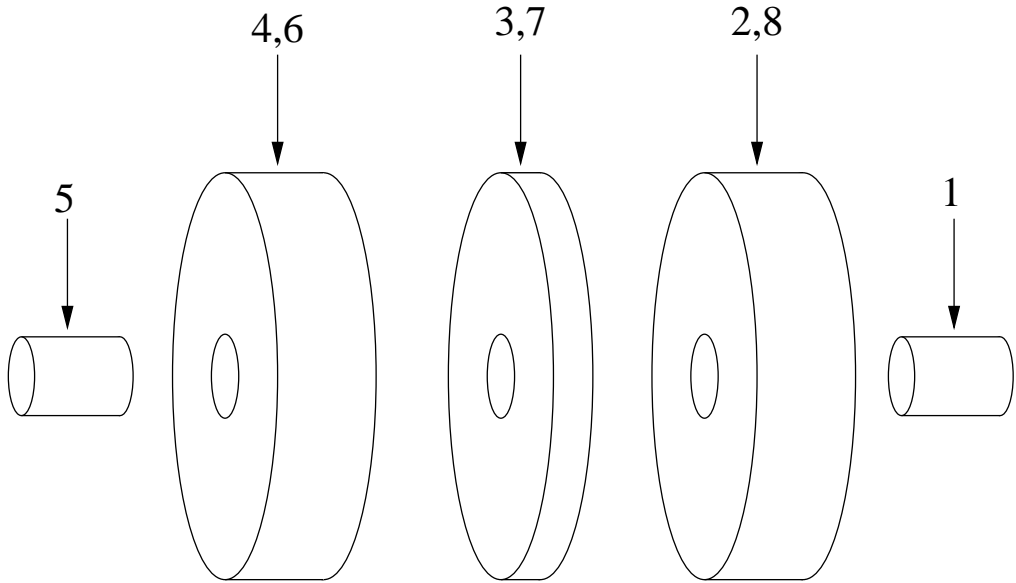}}
\caption{{\sc Domain decomposition for the 3d HRWE Problem.}
Here the inner spherical shell and outer spherical shell, respectively
corresponding to the annuli $H$ and $A$ in Fig.~\ref{domaindecomp},
are not shown. The remaining domains (all lying within the elliptic
region) are shown in an exploded format for emphasis.
} 
\label{domaindecomp3}
\end{figure}

For a 3d problem, the analogous two--center domain is actually composed
of fewer elements than in the equivalent 2d problem. This is one of the 
reasons we believe that a multidomain spectral method is well--suited
for the computations needed in  PSW approximation of binary inspiral. In lieu of a
detailed description of the domain decomposition we envision for 3d
work, we offer the picture in Fig.~\ref{domaindecomp3}.  Note that
cylindrical shell labelled 2,8, for example, corresponds to both
domain 2 and domain 8 in Fig.~\ref{domaindecomp}. Indeed, the 2d
figure could alternatively be viewed as a cross section of a 3d
scenario.

The ten subdomains $H,1-8,A$, along with the symmetry $\psi(-x,-y) =
-\psi(x,y)$, contain the complete information about $\psi$.  Below we
describe a spectral method in which the solution for $\psi$ in each
region is considered as an expansion in terms of appropriate basis
functions. Moreover, we use the term spectral in the truest sense,
since the unknowns we solve for in our numerical approach are in fact
the coefficients of the basis functions. (In a later section, 
while considering the 3d HRWE we will also describe
results based in part on a pseudospectral method in which the unknowns 
are point--values on a spectral grid.)
A crucial feature of our numerical method is to separate the physical
problem into subdomains so that the type changing nature of the HRWE
is then confined within a single subdomain, the outer annulus $A$. The
inner region is then purely elliptical and, despite its two center
topology, this inner part of the manifold is amenable to the standard
spectral techniques associated with elliptic
equations\cite{Pfeifferthesis,PKSTelliptic}.

\subsection{Brief review of mixed--type problems}
In order to place our work in context, we offer some remarks on
type--changing PDE in general. Perhaps the most notable examples arise
in the mathematical description of transonic flow, and in particular
flow over an air foil, a scenario for which subsonic and supersonic
regions are respectively described by elliptic and hyperbolic equations.
The Frankl--Chaplygin equation \cite{Frankl} $K(y) u_{xx} + u_{yy} = 0$
(the choice $K(y) = y$ determines the familiar Tricomi equation
\cite{Tricomi}) serves as a fundamental model
of this behavior, and Morawetz carried out early and fundamental analytic
studies \cite{Morawetz} of it and corresponding first order systems (see
also \cite{Protter}). The first truly successful method for numerically
calculating transonic flow was put forth in the seminal paper \cite{MurmanCole}
by Murman and Cole. They adopted a relaxation--like method along with
type--dependent finite difference stencils. For an account of the impact of
the Murman--Cole method in aerodynamics, and a review of more modern CFD
approaches towards transonic flow, see the historical perspective of
\cite{CaugheyJameson}. While the aerodynamic scenario is the most widely
known, special examples of mixed--type equations arise in fields as diverse
as plasma physics \cite{Otway} and windshield design \cite{Bila}. An
early work most closely related to our own is Chaohao's examination
\cite{Chaohao} of amplifying spiral wave solutions to the 2+1 wave
equation. Numerical methods for mixed--type problems tend to be equation
specific. The Murman--Cole technique, for example, would seem to have
no application for our problem.

Motivated by mixed--type problems, in the late 1950s Friedrichs
initiated a program \cite{Friedrichs} for analyzing a wide class of boundary value
problems based on operators whose symmetric part is positive definite and which
obey certain admissible boundary conditions. Such {\em symmetric positive}
systems include PDE of hyperbolic, elliptic, and mixed type. A lucid history
of the Friedrichs program from both theoretical and numerical perspectives is
given by Jensen \cite{Jensen}. Rather early on, Katsanis developed a numerical
method for solving Friedrichs systems \cite{Katsanis}. Starting with a 
Friedrichs system, he applied Green's theorem in a generic cell, and 
approximated the resulting integral
equation. At the discrete level the approach faithfully mimicked both both the
symmetric positive aspect of the operators as well as admissibility of boundary
conditions, and Katsanis went on to numerically examine the Tricomi problem
\cite{KatsanisTricomi}. Although geometrically flexible, the Katsanis method
is low--order accurate. A finite-difference method for Friedrichs system was
also outlined by Liu \cite{Liu}. Recent work towards numerically solving Friedrichs
systems has drawn on the powerful framework of discontinuous Galerkin methods
\cite{Jensen,EmGuermond}. Although Torre has shown that the 2d HRWE on a
disk can be cast into a first--order Friedrichs system
(and, indeed, the aforementioned work by Chaohao also made connections
with Friedrichs theory), we have nevertheless chosen a classic multidomain
spectral method over a discontinuous Galerkin method. This would seem
appropriate given the relatively simple geometry of our problem and the
expected smoothness of the solutions we seek.

\section{Outer annulus and outer boundary conditions}
\label{sec:outerAnnBC}
We begin by discussing the 2d HRWE on the outer annulus, labeled $A$
in Fig.~\ref{domaindecomp}; with appropriate changes this discussion
will also apply to the 3d HRWE on an outer spherical shell. As the
relevant PDEs are type--changing on them, these are the most interesting
subdomains.  In addition, our discussion here will supply some of the
details of the HRWE and will exhibit its mathematical features,
including its change of type.

In this section, we will describe the radiative outer boundary
conditions on $A$ imposed at a large $r = r_\mathrm{max} \equiv R$.  
We shall also speak of an inner boundary condition for $A$, although 
when $A$ is considered as a subdomain it is not associated with 
a true inner boundary condition. The philosophy here is that we must
understand the relevant boundary value problem on each subdomain to
have all subdomains successfully ``glued'' together.  Inner boundary
conditions on $A$ are imposed at a radial value $r =
r_\mathrm{min}\equiv \varepsilon$\,.  Note that $\varepsilon$ is 
not small. For the decomposition illustrated in Fig.~\ref{domaindecomp}, 
in fact, $\varepsilon$ must be greater than the orbital radius.

\subsection{2d HRWE} The linear 
2d HRWE of Whelan, Krivan, and Price\cite{WheKrivPri} starts with
$-\partial_t^2\psi+\nabla^2\psi=g$ and, in terms of polar coordinates $r,\phi$, the helical
reduction requires that $\psi$ be a function only of $r$ and
$\varphi=\phi-\Omega t$. The 2d HRWE then takes the form
\begin{equation}
\frac{1}{r}
    \frac{\partial}{\partial r}
    \left(r\frac{\partial\psi}{\partial r}
    \right)
  + \frac{1}{r^2}\zeta(r)
    \frac{\partial^2\psi}{\partial\varphi^2} =
g(r,\varphi)\, ,
\label{helical0}
\end{equation}
where $g(r,\varphi)$ is a $\varphi$--periodic source term, $\zeta(r) = 1 - r^2 
\Omega^2$, and the coordinate ranges are
\begin{equation}
\varepsilon \leq r \leq R,\qquad 0 \leq \varphi \leq 2\pi.
\end{equation}
Equation (\ref{helical0}) is clearly elliptic for $r < |\Omega|^{-1}$ and 
hyperbolic for $r > |\Omega|^{-1}$. One boundary value problem would
be to seek solutions to this equation which obey the boundary 
conditions
\begin{equation}
\left.\psi\right|_\varepsilon = \alpha (\varphi),
\qquad
\left.\left(\frac{\partial\psi}{\partial r}
- \Omega \frac{\partial\psi}{\partial \varphi}
+\frac{\psi}{2r}\right)
\right|_{R} = \beta (\varphi),
\label{BCs}
\end{equation}
here with $\varphi$--periodic functions $\alpha (\varphi)$ and $\beta 
(\varphi)$. This boundary value problem for annulus $A$ is equivalent to 
the punctured disk examined by Torre\cite{Torre1}. His proof of the
well--posedness of this problem assumes that the radial endpoint
$\varepsilon$ lies in the elliptic region, as do we by requiring
$\epsilon<\Omega^{-1}$. Torre's proof places no restriction 
on $R$. He allows for $R$ to lie in the hyperbolic region, the
elliptic region, or even on the {\em light cylinder} $r =
|\Omega|^{-1}$. In practice of course, $R$ is large and lies in the
hyperbolic region, as we will assume.

Fourier transformation of (\ref{helical0}) yields
\begin{equation}
\frac{1}{r}
    \frac{\mathrm{d}}{\mathrm{d} r}
    \left(r\frac{\mathrm{d}\hat{\psi}_n}{\partial r}
    \right)
  -\frac{n^2}{r^2} \zeta(r) \hat{\psi}_n =
\hat{g}_n(r)\, ,
\label{helicalft}
\end{equation}
and for the corresponding boundary conditions on the mode 
$\hat{\psi}_n(r)$ we have
\begin{equation}
\hat{\psi}_n(\varepsilon) = \hat{\alpha}_n,  \quad 
\left.\left(
\frac{\mathrm{d}\hat{\psi}_n}{\mathrm{d}r} 
-\mathrm{i}n\Omega \hat{\psi}_n
+ \frac{\hat{\psi}_n}{2r}
\right)\right|_R = \hat{\beta}_n.
\end{equation}
For now we will assume $n \neq 0$, and consider the zero--mode 
case separately later on. As an alternative we may also 
impose exact outgoing--wave boundary conditions, enforced 
$n$--by--$n$ via
\begin{equation}\label{exactBC2d}
\left.\left(
\frac{\mathrm{d}\hat{\psi}_n}{\mathrm{d}r}
-\mathrm{i}n\Omega\hat{\psi}_n
+ \frac{\hat{\psi}_n}{2r}
\right)\right|_R
= \frac{1}{R}\left[n\Omega R \frac{
V'_n(n\Omega R)}{V_n(n\Omega R)}\right]
\hat{\psi}_n,
\end{equation}
where 
\begin{equation}\label{Vnudef}
V_\nu(z) = 
          \sqrt{\frac{\pi z}{2}}
          \exp\big[-\mathrm{i}
          \big(z - {\textstyle\frac{1}{2}} \pi \nu 
        - {\textstyle \frac{1}{4}}\pi\big)\big]
          H^{(+)}_\nu (z)
\end{equation}
is set up to satisfy $V_\nu(z)\sim 1$ as $z \rightarrow\infty$. Here 
$H^{(+)}_\nu(z)$ is the first cylindrical Hankel function. The 
``frequency--domain kernel''
\begin{equation}
n\Omega R \frac{
V'_n (n\Omega R)}{V_n(n\Omega R)}
\end{equation}
can be computed as a continued fraction via Steed's 
algorithm\cite{ThomBarn}. Similar kernels appear in studies of 
radiation boundary conditions for time--domain wave propagation, for 
example in Refs.~\cite{ActaNum,aghsiam,lau04a,HL}. 

\subsection{3d HRWE}\label{subsec:prelim3dHRWE}
For the 3d case we will mostly use the same symbols as for the 2d
case. The 3d HRWE of Andrade {\em et al.}\cite{Andradeetal} is
\begin{equation}
\frac{1}{r^2}
    \frac{\partial}{\partial r}
    \left(r^2\frac{\partial \psi}{\partial r}
    \right)
  + \left(\frac{\Delta_{S^2}}{r^2}
  - \Omega^2
    \frac{\partial^2}{\partial\varphi^2}\right) \psi
  = g(r,\theta,\varphi)\, ,
\label{helical30}
\end{equation}
where $g(r,\theta,\varphi)$ is again a $\varphi$--periodic source term,
$\Delta_{S^2}$ is the unit two--sphere Laplacian, and the coordinate 
ranges now are
\begin{equation}
\varepsilon \leq r \leq R,\quad 0\leq \theta\leq\pi,\quad 0 \leq
\varphi \leq 2\pi.
\end{equation}
Equation (\ref{helical30}) is elliptic for $r\sin\theta < |\Omega|^{-1}$ 
and hyperbolic for $r\sin\theta > |\Omega|^{-1}$. We seek solutions
to this equation which obey the boundary conditions
\begin{equation}
\left. 
\psi\right|_\varepsilon = \alpha(\theta,\phi)\, \qquad
\left.\left(\frac{\partial \psi}{\partial r}
- \Omega \frac{\partial \psi}{\partial \varphi}
+ \frac{\psi}{r}\right)
\right|_{R} = \beta(\theta,\phi).
\label{3BCs}
\end{equation}
We consider the exact 
outgoing--wave conditions below. As before, we will assume
that $\varepsilon < |\Omega|^{-1}$, and explain why in a moment. 

Spherical harmonic transformation of (\ref{helical30}) yields
\begin{equation}
\frac{1}{r^2}
    \frac{\mathrm{d}}{\mathrm{d} r}
    \left(r^2\frac{\mathrm{d}
\hat{\psi}_{\ell m}}{\mathrm{d} r}
    \right)
  -\left[\frac{\ell(\ell+1)}{r^2} 
  - m^2\Omega^2\right]\hat{\psi}_{\ell m}
  = \hat{g}_{\ell m}(r)\, ,
\label{helical3d}
\end{equation}
and we will now take
\begin{equation}
\hat{\psi}_{\ell m}(\varepsilon) = \hat{\alpha}_{\ell m},  \quad
\left.\left(
\frac{\mathrm{d}\hat{\psi}_{\ell m}}{\mathrm{d}r}
-\mathrm{i}m\Omega \hat{\psi}_{\ell 
m}
+ \frac{\hat{\psi}_{\ell m}}{r}
\right)\right|_R = \hat{\beta}_{\ell m}
\end{equation}
as the corresponding boundary conditions. We will assume $m \neq 0$, 
and consider the $m = 0$ case separately later on. As an alternative we 
may also impose exact outgoing--wave boundary conditions. Enforced 
mode--by--mode, they have a form similar to Eq.~(\ref{exactBC2d}),
\begin{equation}
\left.\left(
\frac{\mathrm{d}\hat{\psi}_{\ell m}}{\mathrm{d}r}
-\mathrm{i}m\Omega\hat{\psi}_{\ell m}
+ \frac{\hat{\psi}_{\ell m}}{r}
\right)\right|_R
= \frac{1}{R}\left[m\Omega R \frac{
V'_{\ell+1/2}(m\Omega R)}{V_{\ell+1/2}(m\Omega R)}\right]
\hat{\psi}_{\ell m}\,.
\end{equation}

To cast the radial equation stemming from the 3d HRWE 
into a form which resembles the radial equation stemming from the 
2d HRWE, we substitute $\hat{\psi}_{\ell m} = \hat{\xi}_{\ell m}/\sqrt{r}$
in (\ref{helical3d}), thereby finding
\begin{equation}
\frac{1}{r}
    \frac{\mathrm{d}}{\mathrm{d} r}
    \left(r\frac{\mathrm{d}
\hat{\xi}_{\ell m}}{\mathrm{d} r}
    \right)
  - \frac{(\ell+{\textstyle\frac{1}{2}})^2}{r^2}
    \zeta(r)\hat{\xi}_{\ell m}
  = \sqrt{r}\hat{g}_{\ell m}(r)\, ,
\label{helical3to2}
\end{equation}
where 
\begin{equation}
\zeta(r) = 1 - \frac{4 m^2\Omega^2 r^2}{(2\ell+1)^2}.
\end{equation}
In terms of $\omega_{\ell m} = 2 m\Omega/(2\ell+1) \neq 0$, 
clearly something special occurs for (\ref{helical3to2}) when 
$r = |\omega_{\ell m}|^{-1}$, here with $\ell, m \neq 0$. Note 
that $|\omega_{\ell m}| \leq |\Omega|$, so that 
$|\Omega|^{-1} \leq |\omega_{\ell m}|^{-1}$. For this reason, 
we have required $\varepsilon < |\Omega|^{-1}$ above, ensuring
that $r = \varepsilon$ lies within each mode's individual 
``elliptic region.''

\subsection{General form for the outer boundary conditions}
From Eq.~(\ref{helicalft}) we  pass to a trigonometric rather 
than exponential representation of 
the transform $\hat{\psi}_n(r)$ by writing
\begin{eqnarray}
\lefteqn{\hat{\psi}_{n}^*(r)\exp(-\mathrm{i}n\varphi)
+ \hat{\psi}_n(r)\exp(\mathrm{i}n\varphi)} & &
\label{defofuandw}
\\
& = & 
\big[\hat{\psi}_{n}(r) + \hat{\psi}^*_n(r)\big] 
\cos(n\varphi)
+\mathrm{i}\big[\hat{\psi}_{n}(r) - \hat{\psi}^*_n(r)\big]
\sin(n\varphi)
\nonumber \\
& = & u_n(r)\cos(n\varphi) + w_n(r)\sin(n\varphi),
\nonumber
\end{eqnarray}
with a similar splitting possible for $\hat{\psi}_{\ell m}(r)$,
or $\hat{\xi}_{\ell m}(r)$ if preferred.

At $r = R$ all boundary conditions, whether exact
or some incarnation of Sommerfeld, may be expressed as follows:
\begin{equation}
Rw_n'(R) + p u_n(R) + q w_n(R) = 0,\quad
Ru_n'(R) - p w_n(R)
+ q u_n(R) = 0.
\label{rmaxuBCs}
\end{equation}
We list the possibilities considered so far in 
Table \ref{tab:table1}. Other choices of the form (\ref{rmaxuBCs}) are of 
course possible. For the exact conditions listed in Table \ref{tab:table1}, 
we have made use of the notation
\begin{equation}
 v_\nu(z) = z \frac{
V'_\nu(z)}{V_\nu(z)},
\end{equation}
with the obvious notation for real and imaginary parts. 
\begin{table}
\centering
\begin{tabular}{|l|c|c|}
\hline
Boundary condition & $p$ & $q$ \\
\hline
Sommerfeld on $\hat{\psi}_n$ & $n\Omega R$ & ${\textstyle\frac{1}{2}}$
\\
Sommerfeld on $\hat{\psi}_{\ell m}$ & $m\Omega R$ & $1$
\\
Exact on $\hat{\psi}_n$ &
$n\Omega R +\mathrm{Im}  v_n(n\Omega R)$ &
${\textstyle\frac{1}{2}} - \mathrm{Re}  v_n(n\Omega R)$ \\
Exact on $\hat{\psi}_{\ell m}$ &
$m\Omega R +\mathrm{Im}  v_{\ell+1/2}(m\Omega R)$
&
$1 - \mathrm{Re}  v_{\ell+1/2}(m\Omega R)$
\\
\hline
\end{tabular}
\caption{\label{tab:table1}{\sc Outer boundary
conditions for the 2d and 3d HRWE.}}
\end{table}

Turning now to the rather more delicate zero--modes ($n = 0$ for
2d and $m = 0$ for 3d), we note that for $z\rightarrow0$ the Hankel
function $H_\nu^{(+)}(z)$ is proportional to $z^{-\nu}$ for $\nu\neq0$
and to $\log{z}$ for $\nu=0$, so that in either case  
\begin{equation}
z H^{(+)}_\nu{}'(z)/H^{(+)}_\nu (z)\rightarrow -\nu\,,
\end{equation}
and hence from the the definition in
Eq.~(\ref{Vnudef}) we have 
\begin{equation}
 v_\nu(0) = {\textstyle\frac{1}{2}} - \nu
\label{myveeresult}\,.
\end{equation}
From this our key result follows.
For the $n = 0$ homogeneous case, Eq.~(\ref{helicalft}) has solutions 
$\hat{\psi}_0(r) = c$ or $\hat{\psi}_0(r) \propto \log r$. With $ v_0(0) 
= {\textstyle\frac{1}{2}}$, we find $p = 0 = q$ for the exact 
outgoing conditions as stated in (\ref{rmaxuBCs}). So this boundary 
condition rules out the $\log r$ solution. For the $m = 0$ homogeneous 
case, Eq.~(\ref{helical3d}) has solutions $\hat{\psi}_{\ell 0} \propto
r^\ell$ and $\hat{\psi}_{\ell 0} \propto r^{-(\ell+1)}$. Now 
$ v_{\ell+1/2}(0) = -\ell$, and $p = 0$, $q = \ell + 1$ for the
exact outgoing boundary condition. In this 
case the boundary condition rules out the $r^\ell$ solution. 

\section{Sparse spectral approximation of the 2d HRWE}
\label{sec:sparse}

This section describes how we use spectral methods to numerically
solve the 2d HRWE on the two center domain shown in 
Fig.~\ref{domaindecomp}. 
Our method will exploit the spectral {\em integration 
preconditioning} (IPC) proposed by Coutsias, Hagstrom, Hesthaven, 
and Torres\cite{CHHT}, a general--purpose spectral method for
solving ODEs and PDEs. Developing IPC
for general orthogonal polynomials, they mostly considered it
in the context of ODEs. In this context they presented a
detailed theoretical analysis of both conditioning
and convergence (see also \cite{CHT}). Although they also carefully outlined
how to apply the method in higher dimensions (with several 
illuminating two--dimensional examples), 
they did not explore conditioning issues for PDEs.

We shall follow the IPC approach for the following reasons. Foremost,
it is a direct recipe for applying spectral methods in a PDE setting,
affords a straightforward way of treating both
boundary conditions and the ``gluing'' of subdomains. Moreover, IPC
allows us simultaneously to achieve a sparse banded matrix
representation of the 2d HRWE on all subdomains. While we question
whether a fully 3d multidomain PSW problem can be treated exclusively
with spectral methods based on IPC (further comments to follow), we
believe the method is well suited for handling the 2d HRWE on the outer
annulus $A$ (or the 3d HRWE on an outer spherical shell). The
mixed--type boundary value problem on this subdomain introduces
several novel features, and much of our analysis below centers on
associated conditioning issues.  In short, the IPC approach offers us
a quick and easy way to test the fundamental idea of using a
multidomain spectral method to solve the 2d HRWE, and has promise 
for at least one key aspect of true 3d PSW problems.

In following the IPC approach we are able to explicitly form the
matrix which represents the 2d HRWE on the two center domain, and we
subsequently use Gaussian elimination (as embodied by the {\tt netlib}
routine {\tt dgesv}, and sometimes also {\tt dgesvx}) to invert the 
resulting system. A discussion of the structure of the matrix serves to 
sharpen our analytic
understanding of the linear systems we are dealing with. For 
3d problems, however, speed and memory considerations
make it impractical to form and solve the full matrix. 
For the 3d model we consider in the
conclusion and for envisioned future 3d work, we have used, and 
plan to continue with 
the Krylov--based method GMRES, in which only the specification of a
matrix--vector multiply need be implemented. For a Krylov method
preconditioning strategies are often necessary to avoid stagnation of
the iterative solver.

Despite the name of the method, and despite its intuitive appeal, it
is not guaranteed that IPC, especially for rectangular domains,
actually improves conditioning (with respect to the problem of 
matrix inversion). If this is an issue in two dimensions, it is sure
to be even more problematic in three
dimensions. Since we cannot guarantee that our implementation of IPC
will actually improve conditioning, it would be more appropriate to
call the method a ``sparse formulation.'' To refer to the very
specific IPC technique it will be convenient, however, for us to
continue to use the term ``preconditioning.'' For
simplicity, and to have a uniform treatment, we will use the IPC method
for all our subdomains in the 2d model. It should be understood that
this uniformity is not necessary. We could treat the elliptic region
via a different method (spectral or pseudospectral). In any case, as
we verify with numerical experiments, the method yields impressive global 
accuracy in solving the 2d HRWE on our two center domain.

\subsection{Basic formulas for Chebyshev polynomials}
\label{basicChebyForms}
Although Ref.~\cite{CHHT} considered general orthogonal polynomials, 
we work solely with Chebyshev polynomials, 
a classical system of orthogonal polynomials with many useful 
properties and applications\cite{Rivlin,Beckman}. 
Here we collect only those properties 
relevant for our discussion, mostly following \cite{CHHT,H,WKC}. 
The degree--$n$ Chebyshev polynomial $T_n(\xi)$ 
is defined by $T_n(\xi)  = \cos(n\arccos(\xi))$ for $-1 \leq \xi 
\leq 1$, showing that we may consistently set $T_{-n}(\xi)
= T_n(\xi)$. In our application $\xi$ depends on one of the coordinates, 
say $\xi(r) = (2r - r_\mathrm{max} - r_\mathrm{min})/(r_\mathrm{max} - 
r_\mathrm{min})$ in an outer spherical shell or annulus. 
The $T_n(\xi)$ are solutions to a {\em singular} 
Sturm--Liouville problem, and therefore particularly suited for 
approximating solutions to differential equations on 
$[-1,1]$ with a wide class of boundary conditions.
The $T_n(\xi)$ are orthogonal on the interval $[-1,1]$ 
with respect to the weight function $(1-\xi^2)^{-1/2}$.

We have $T_0(\xi) = 1$, $T_1(\xi) = \xi$, and the following 
identity:
\begin{equation}
2\xi T_n(\xi) = T_{n+1}(\xi) + T_{n-1}(\xi).
\label{multbyx}
\end{equation}
If we denote by $\mathbf{T}(\xi)$ the row vector 
$[T_0(\xi), T_1(\xi), T_2(\xi), \cdots]$, then (\ref{multbyx})
may be expressed as
\begin{equation}
\xi\mathbf{T}(\xi) = \mathbf{T}(\xi)A,
\end{equation}
where\footnote{For now we proceed with infinite--size 
matrices. However, numerical approximation necessarily 
entails suitable truncations. Our practice will be to use the same 
symbols for such truncations, and alert the reader whenever the 
viewpoint shifts.}
\begin{equation}\label{AT0}
A = \left[
\begin{array}{ccccccc}
0 & \frac{1}{2} & 0 & 0 & 0 & 0 & \cdots\\
1 & 0 & \frac{1}{2} & 0 & 0 & 0 & \cdots\\
0 & \frac{1}{2} & 0 & \frac{1}{2} & 0 & 0 & \cdots\\
0 & 0 & \frac{1}{2} & 0 & \frac{1}{2} & 0 & \cdots\\
0 & 0 & 0 & \frac{1}{2} & 0 & \frac{1}{2} & \cdots\\
0 & 0 & 0 & 0 & \frac{1}{2} & 0 & \cdots\\
\vdots & \vdots & \vdots & \vdots & \vdots & \vdots & \ddots
\end{array}
\right].
\end{equation}
Note that our convention here is to label matrix rows and
columns starting from 0 rather than 1. Since $A$ is tridiagonal,
the matrix $A^m$,  representing left multiplication by $\xi^m$
through the formula $\xi^m\mathbf{T}(\xi) = \mathbf{T}(\xi)A^m$, 
has bandwidth $2m+1$. Moreover, banded matrices $p(A)$ similarly 
correspond to multiplication by any polynomial $p(\xi)$. The most
important special case, is the matrix $A_{T_m}\equiv T_m(A)$ of 
bandwidth $2m+1$ corresponding to multiplication by the Chebyshev 
polynomial $T_m(\xi)$. For $m=1$, the matrix 
$A=A_{T_1}$ is given by Eq.~(\ref{AT0}). More generally the
entries of $A_{T_m}$ may be 
gathered from the identity\cite{CHHT,Rivlin}
\begin{equation}
2 T_m(\xi) T_n(\xi) = T_{m+n}(\xi) + T_{|m-n|}(\xi)\,.
\label{TxTequalsT}
\end{equation}

Next we consider the formal expansion 
\begin{equation}
u(\xi) = \sum_{n=0}^\infty 
\widetilde{u}_n T_n(\xi),
\label{basicexp}
\end{equation}
where, owing to the orthogonality of the Chebyshev basis, the
expansion coefficients have the analytic values\cite{Rivlin}
\begin{equation}
\widetilde{u}_n = \kappa_n
\int_{-1}^{1}u(\xi)T_n(\xi)\frac{d\xi}{\sqrt{1-\xi^2}}\,.
\label{continuousun}
\end{equation}
The prefactor $\kappa_n$  is $1/\pi$ for $n = 0$ and $2/\pi$ for $n>0$ .
We denote by $\widetilde{\mathbf{u}}$ the column vector of expansion 
coefficients $[\widetilde{u}_0, \widetilde{u}_{1}, \widetilde{u}_{2}, 
\cdots]^t$. In the vector space of expansion coefficients 
multiplication by $\xi$ is effected through multiplication 
by $A$, that is to say $\widetilde{\mathbf{u}}{}^\xi 
\equiv A\widetilde{\mathbf{u}}$ represents the 
expansion coefficients 
$[\widetilde{u}_0^\xi, \widetilde{u}_1^\xi,\widetilde{u}_{2}^\xi,
\widetilde{u}_{3}^\xi,\cdots]^t$ for 
$u^\xi(\xi) \equiv \xi u(\xi)$. Similarly, in the space of expansion
coefficients, multiplication by any polynomial $p(\xi)$ may be represented
via left multiplication by $p(A)$.

Another well--known formula\cite{Rivlin,CHHT,H}
\begin{equation}
T_n(\xi) = \frac{T'_{n+1}(\xi)}{2(n+1)}
        -\frac{T'_{n-1}(\xi)}{2(n-1)},
\label{TfromTp}
\end{equation}
yields another matrix identity
\begin{equation}
\mathbf{T}(\xi) = \mathbf{T}'(\xi) B_{[1]}, 
\end{equation}
in terms of an {\em integration matrix}
\begin{equation}
B_{[1]} = \left[
\begin{array}{ccccccc}
0 & 0 & 0 & 0 & 0 & 0 & \cdots\\
1 & 0 & -\frac{1}{2} & 0 & 0 & 0 & \cdots\\
0 & \frac{1}{4} & 0 & -\frac{1}{4} & 0 & 0 & \cdots\\
0 &  0  & \frac{1}{6} & 0 & -\frac{1}{6} & 0  & \cdots\\
0 & 0  & 0  & \frac{1}{8} & 0 & -\frac{1}{8} & \cdots\\
0 & 0 & 0  & 0  & \frac{1}{10} & 0 & \cdots\\
\vdots & \vdots  & \vdots  & \vdots  & \vdots & \vdots & \ddots
\end{array}
\right].
\end{equation}
Here the subscript $[1]$ merely emphasizes that the first row of 
$B_{[1]}$ consists only of zeros. Adopting a first row of zeros
is a choice, permissible since $T_0'(\xi) = 0$. Free, or open, 
rows of zeros within integration matrices is a central feature of
the spectral method presented in \cite{CHHT}, and following that
reference we will eventually exploit such freedom to enforce
boundary conditions. In matrix form the formula
\begin{equation}
T_n(\xi) = \frac{T''_{n+2}(\xi)}{4(n+1)(n+2)}
       -\frac{T''_{n}(\xi)}{2(n^2-1)}
       +\frac{T''_{n-2}(\xi)}{4(n-1)(n-2)}
\label{TfromTpp}
\end{equation}
reads
\begin{equation}
\mathbf{T}(\xi) = \mathbf{T}''(\xi) B^2_{[2]},
\end{equation}
where now
\begin{equation}
B^2_{[2]} = \left[
\begin{array}{ccccccc}
0 & 0 & 0 & 0 & 0 & 0 & \cdots\\
0 & 0 & 0 & 0 & 0 & 0 & \cdots\\
\frac{1}{4} & 0 & -\frac{1}{6} & 0 & \frac{1}{24} & 0 & \cdots\\
0 & \frac{1}{24}  & 0 & -\frac{1}{16} & 0  & \frac{1}{48} & \cdots\\
0 & 0  & \frac{1}{48}  & 0 & -\frac{1}{30} & 0 & \cdots\\
0 & 0 & 0  & \frac{1}{80}  & 0 & -\frac{1}{48} & \cdots\\
\vdots & \vdots  & \vdots  & \vdots  & \vdots & \vdots & \ddots
\end{array}
\right].
\end{equation}
Single and double integration of $\widetilde{\mathbf{u}}$ then
correspond to $B_{[1]}\widetilde{\mathbf{u}}$ and 
$B^2_{[2]}\widetilde{\mathbf{u}}$.

Consider the matrices $D$ and $D^2$ corresponding to differentiation in the 
Chebyshev basis. The entries of these matrices stem from recursive use of 
(\ref{TfromTp}) for $D$ and (\ref{TfromTpp}) for $D^2$. Although we will not 
require the precise forms of these matrices, we nevertheless list them here 
for completeness:
\begin{equation}
D =  \left[
\begin{array}{cccccccccc}
0 & 1 & 0 & 3 & 0 & 5 & 0 & 7 & 0 & \cdots\\ 
0 & 0 & 4 & 0 & 8 & 0 & 12 & 0 & 16 & \cdots\\
0 & 0 & 0 & 6 & 0 & 10 & 0 & 14 & 0 & \cdots\\
0 & 0 & 0 & 0 & 8 & 0 & 12 & 0 & 16 & \cdots\\
0 & 0 & 0 & 0 & 0 & 10 & 0 & 14 & 0 & \cdots\\
0 & 0 & 0 & 0 & 0 & 0 & 12 & 0 & 16 & \cdots\\
0 & 0 & 0 & 0 & 0 & 0 & 0  & 14 & 0 & \cdots\\
0 & 0 & 0 & 0 & 0 & 0 & 0  & 0 & 16 & \cdots\\
\vdots & \vdots & \vdots & \vdots & \vdots & \vdots & \vdots & \vdots &  \vdots & \ddots
\end{array}
\right],
\end{equation}
\begin{equation}
D^2 =  \left[
\begin{array}{cccccccccc}
0 & 0 & 4 & 0  & 32 & 0   & 108 &  0   &  256 & \cdots\\
0 & 0 & 0 & 24 & 0  & 120 & 0   &  336 &  0   & \cdots\\
0 & 0 & 0 & 0  & 48 & 0   & 192 &  0   &  480 & \cdots\\
0 & 0 & 0 & 0  & 0  & 80  & 0   &  280 &  0   & \cdots\\
0 & 0 & 0 & 0  & 0  & 0   & 120 &  0   &  384 & \cdots\\
0 & 0 & 0 & 0  & 0  & 0   & 0   &  168 &  0   & \cdots\\
0 & 0 & 0 & 0  & 0  & 0   & 0   &  0   &  224 & \cdots\\
0 & 0 & 0 & 0  & 0  & 0   & 0   &  0   &  0   & \cdots\\
0 & 0 & 0 & 0  & 0  & 0   & 0   &  0   &  0   & \cdots\\
\vdots & \vdots & \vdots & \vdots & \vdots & \vdots & \vdots & \vdots &  \vdots & \ddots
\end{array}
\right].
\end{equation}
The matrices $D$ and $D^2$ obey
\begin{equation}
\mathbf{T}'(\xi) = \mathbf{T}(\xi)D,\qquad 
\mathbf{T}''(\xi) = \mathbf{T}(\xi)D^2.
\end{equation}
Notice that the first column of $D$ and the first two columns of $D^2$ are 
(necessarily) zero. We might accordingly represent these matrices as 
$D_{\langle 1\rangle}$ and $D_{\langle 2\rangle}$ to highlight these facts, 
but will not employ this notation. Whereas the matrices
$B_{[1]}$ and $B^2_{[1]}$ are banded and sparse, $D$ and $D^2$ are clearly
upper triangular and dense.
To render the linear systems we encounter 
in banded form, we will exploit the identities,\cite{CHHT}
\begin{equation}
B_{[1]}D = I_{[1]},\quad B^2_{[2]}D^2 = I_{[2]}, \quad B^2_{[2]}D 
= B_{[2]}.
\label{BDequalsI}
\end{equation} 
Here the notation $I_{[1]}$ means the identity matrix with the first row 
set to zero; $I_{[2]}$ means the first two rows are zero.
We stress the crucial point that, for the ordering of matrix products
in Eq.~(\ref{BDequalsI}),
the identities (\ref{BDequalsI}) also hold for $(N+1)$--by--$(N+1)$ 
truncations of $B_{[1]}$, $B^2_{[2]}$, $D$, and $D^2$. 

In order to enforce boundary conditions we introduce both
Dirichlet and Neumann vectors (radiation boundary conditions
require a bit more thought and will be put off until later).
The Dirichlet vectors are
\begin{eqnarray}
& & \delta^{+} =  \big[1,1,1,1\dots]
= \big[T_0(1),T_1(1),T_2(1),T_3(1),\cdots\big]
\\
& & \delta^{-} = \big[1,-1,1,-1,\dots]
=  \big[T_0(-1),T_1(-1),T_2(-1),T_3(-1),\cdots\big],
\end{eqnarray}
while the Neumann vectors are
\begin{eqnarray}
& & \nu^{+} =  \big[0,1,4,9,\dots]
= \big[T_0'(1),T_1'(1),T_2'(1),T_3'(1),\cdots\big]
\\
& & \nu^{-} = \big[1,-1,4,-9,\dots]
= \big[T'_0(-1),T'_1(-1),T'_2(-1),T'_3(-1),\cdots\big]\label{numinus}.
\end{eqnarray}
The boundary conditions we encounter are easily expressed in terms of these 
vectors. For example, $\delta^{-} \cdot \widetilde{\mathbf{u}} = 0$ is a 
homogeneous Dirichlet boundary condition at the left endpoint, and $\nu^{+} \cdot 
\widetilde{\mathbf{u}} = 0$ is a homogeneous Neumann boundary condition at the 
right endpoint.

We conclude this preliminary subsection by addressing a few practical
matters.  First, our discussion so far assumes the standard interval
$[-1,1]$. Modifications, albeit trivial ones, of the matrices $A$, 
$B_{[1]}$, $B^2_{[2]}$ and vectors $\nu^\pm$ are needed for different
intervals. For example, if we use the $T_n(\xi(r))$ to approximate
functions on $r\in [r_\mathrm{min},r_\mathrm{max}]$, then to
represent double integration that corresponds to the physical
interval we must send $B^2_{[2]} \rightarrow 0.25(r_\mathrm{max} -
r_\mathrm{min})^2 B^2_{[2]}$.  Similar scalings must be made for $A$,
$B_{[1]}$, and $\nu^\pm$. However, {\em we shall retain the same symbols
for matrices which are so scaled}. Second, although we may
theoretically consider an infinite expansion (\ref{basicexp}) for a
given function $u$ (perhaps a solution to a differential equation),
numerical applications entail a suitable truncation $\mathcal{P}_N
u(\xi) = \sum_{n=0}^N \widetilde{u}_n T_n(\xi)$, where $\mathcal{P}_N$
represents the truncation operator. Moreover, most applications do not
work directly with the analytic expansion coefficients
(\ref{continuousun}). Rather, one typically approximates the integral
in (\ref{continuousun}) via the quadrature rule stemming from
Chebyshev--Gauss--Lobatto nodes (see, for example, \cite{H}). This
process introduces aliasing errors, and results in discrete expansion
coefficients $\widetilde{u}{}_n^\mathrm{discrete}$. A more theoretical
treatment would, when appropriate, draw a careful distinction between
the analytic and discrete expansion coefficients, but this issue is
not of primary importance to us.

\subsection{Sparse formulation on rectangular domains}

In terms of co--moving Cartesian coordinates, 
\begin{equation}
x = r \cos\varphi = r\cos(\phi-\Omega t),\qquad
y = r\sin\varphi = r\sin(\phi-\Omega t),
\end{equation}
let the scalar field $\psi(x,y)$ obey the inhomogeneous 
2d HRWE
\begin{equation}
\left[ \partial_x^2 +
       \partial_y^2 -
\Omega^2\big(x\partial_y  - y\partial_x\big)^2
\right]\psi = g
\label{rectHRWE}
\end{equation}
on a rectangle $R = [x_\mathrm{min},x_\mathrm{max}]
\times [y_\mathrm{min},y_\mathrm{max}]$. We have also worked 
with the shifted equation resulting from the mapping $x\rightarrow 
x-c$, $y\rightarrow y-d$, where $(c,d)$ is the center of the 
rectangle. For simplicity, we do not consider a shift 
here. Via repeated use of the Leibniz rule, we write 
(\ref{rectHRWE}) as  
\begin{equation}
\left[
  \partial_x^2 (1 - \Omega^2 y^2)
+ \partial_y^2 (1 - \Omega^2 x^2)
- \Omega^2\big( \partial_x x
              + \partial_y y
             -2 \partial_x\partial_y x y\big)
              \right]\psi = g,
\label{rectHRWE2}
\end{equation}
in preparation for our sparse--matrix spectral approximation. 

At the theoretical level, we may represent our solution in terms of a 
double Chebyshev expansion,
\begin{equation}
\psi(x,y) =
\sum_{n=0}^\infty
\sum_{m=0}^\infty
\widetilde{\psi}_{nm}
T_n(\xi(x))T_m(\eta(y)),
\end{equation}
where $(\xi(x),\eta(y))$ is a mapping of our rectangle $R$ onto the unit
rectangle $[-1,1]\times [-1,1]$. To obtain the system of equations we 
solve numerically, we consider the truncated series 
\begin{equation}
\mathcal{P}_{N,M}\psi(x,y) = \sum_{n=0}^N\sum_{m=0}^M
\widetilde{\psi}_{nm} T_n(\xi(x))T_m(\eta(y)).
\end{equation}
We represent the finite collection of expansion coefficients as a 1--vector 
\begin{equation}\label{ordering}
\widetilde{\boldsymbol{\psi}} =
\big(
\widetilde{\psi}_{00},
\widetilde{\psi}_{01},
\cdots,
\widetilde{\psi}_{0M},
\widetilde{\psi}_{10},
\widetilde{\psi}_{11},
\cdots,
\widetilde{\psi}_{1M},
\cdots,
\widetilde{\psi}_{N0},
\widetilde{\psi}_{N1},
\cdots
\widetilde{\psi}_{NM}\big)^t,
\end{equation}
so that the components $\widetilde{\boldsymbol{\psi}}(k) =
(\widetilde{\boldsymbol{\psi}})_k$ are determined by the direct
product representation
\begin{equation}
\widetilde{\boldsymbol{\psi}}(n(M+1)+m) = \widetilde{\psi}_{nm}.
\end{equation}
We then consider the approximation of (\ref{rectHRWE2}) in terms of
$\widetilde{\boldsymbol{\psi}}$ and suitable truncations of spectral
differentiation matrices,
\begin{align}
& \big[
  D_x^2 \otimes (I_y - \Omega^2 A_y^2)
+ (I_x - \Omega^2 A_x^2) \otimes D_y^2
\nonumber \\
& - \Omega^2\big(D_xA_x \otimes I_y
              + I_x \otimes D_yA_y
             -2 D_x A_x \otimes D_y A_y\big)
              \big] \widetilde{\boldsymbol{\psi}} =
              \widetilde{\mathbf{g}}.
\label{Rspecform}
\end{align}
Here $D_x$ represents the $(N+1)$--by--$(N+1)$ differentiation matrix 
in the Chebyshev basis $T_n(\xi(x))$, and $A_x$ represents the 
$(N+1)$--by--$(N+1)$ matrix corresponding to multiplication by $x$, 
with similar statements for $D_y$ and $A_y$ which are both 
$(M+1)$--by--$(M+1)$. As mentioned at the end of the last subsection,
to obtain these matrices appropriate scaling factors must be included 
with the straightforward truncations of the infinite--size matrices 
listed above. By the definition of the Kronecker direct
product, we have, for example, that
\begin{equation}
(D_x^2 \otimes A_y^2)(n(M+1)+m,k(M+1)+p) = (D_x^2)(n,k) (A_y^2)(m,p),
\end{equation}
in the {\sc Matlab} notation $C(i,k)$ for entries of a matrix $C_{ik}$.

To achieve a sparse and banded representation of the approximation, 
we multiply (\ref{Rspecform}) by $\mathcal{B} = B^2_{x[2]} \otimes 
B^2_{y[2]}$ and exploit the identities (\ref{BDequalsI}), thereby
reaching
\begin{align}\label{Kronform}
\left[
I_{x[2]}\otimes B^2_{y[2]} (I_y - \Omega^2 A_y^2)
+  
B^2_{x[2]}(I_x - \Omega^2 A_x^2)\otimes I_{y[2]}
-\Omega^2\big(B_{x[2]}A_x \otimes B^2_{y[2]} \right. 
&
\nonumber \\
\left.              
+
B^2_{x[2]}\otimes B_{y[2]}A_y
             -2
B_{x[2]}A_x \otimes B_{y[2]}A_y\big)
              \right] \widetilde{\boldsymbol{\psi}} =
B^2_{x[2]}\otimes B^2_{y[2]}
\widetilde{\mathbf{g}}\,. &
\end{align}
The coefficient matrix of the system is then
\begin{displaymath}
 \left[
\begin{array}{cccccccc}
0 & 0 & 0 & 0 & 0 & 0 & \cdots & 0\\
0 & 0 & 0 & 0 & 0 & 0 & \cdots & 0\\
\frac{I_{[2]}}{4} & 0 & -\frac{I_{[2]}}{6} & 0 & \frac{I_{[2]}}{24} & 0 & \cdots & 0\\
0 & \frac{I_{[2]}}{24}  & 0 & -\frac{I_{[2]}}{16} & 0  & \frac{I_{[2]}}{48} & \cdots & 0\\
0 & 0  & \frac{I_{[2]}}{48}  & 0 & -\frac{I_{[2]}}{30} & 0 & \cdots & 0\\
0 & 0 & 0  & \frac{I_{[2]}}{80}  & 0 & -\frac{I_{[2]}}{48} & \cdots & 0\\
\vdots & \vdots  & \vdots  & \vdots  & \vdots & \vdots & \ddots & \vdots \\
0 & 0 & 0 & 0 & 0 & \frac{I_{[2]}}{4N(N-1)} & 0 & -\frac{I_{[2]}}{2(N^2-1)}
\end{array}
\right] +
\end{displaymath}
\begin{displaymath}
\left[\begin{array}{cccccccc}
 0 & 0 & 0 & 0 & 0 & 0 & \cdots & 0\\
 0 & 0 & 0 & 0 & 0 & 0 & \cdots & 0\\
0 &
0 &
B^2_{[2]} &
0 &
0 & 0 &
\cdots & 0\\
 0 & 0  & 0 &
B^2_{[2]} & 0 &
0 & \cdots & 0\\
0 & 0 & 0  & 0 &
B^2_{[2]} & 0 & \cdots & 0 \\
0 & 0 & 0 & 0  & 0 &
B^2_{[2]} & \cdots & 0 \\
\vdots & \vdots & \vdots & \vdots & \vdots & \vdots &
\ddots & \vdots \\
 0 & 0 & 0 & 0 & 0 & 0 & \cdots & B^2_{[2]} 
\end{array}\right] - \Omega^2
\left[\cdots\right]
\end{displaymath}
where we have only explicitly displayed the $\Omega$--independent
contribution to the matrix stemming from the preconditioned
Laplacian. This Laplacian contribution exhibits the main features of
the overall matrix. Namely, that it is banded, sparse, and contains
free rows of zeros. Each entry in either of the matrices above is
itself an $(M+1)$--by--$(M+1)$ matrix. Indeed, the $I_{[2]}$ and
$B^2_{[2]}$ matrices are $I_{y[2]}$ and $B^2_{y[2]}$ in the notation
of Eq.~(\ref{Kronform}), and in the $y$ dimension the truncation is
$m=0,\cdots,M$. Overall, the coefficient matrix has $(N+1)^2$ such
blocks. The first two block--rows in the matrices displayed above are 
empty, giving $2(M+1)$ free rows of zeros. In each of the remaining 
$N-1$ block--rows the first two rows are empty, giving us another 
$2(N-1)$ zero rows. Therefore, we have a total of $2(N+M)$ such 
zero rows at our disposal.

In the free zero rows we insert the ``$\tau$--conditions'' \cite{CHHT},
that is, the boundary conditions. To illustrate, we here assume
Dirichlet conditions
$\psi(x_\mathrm{min},y) = f^{-}(y)$, $\psi(x_\mathrm{max},y) = 
f^{+}(y)$,
$\psi(x,y_\mathrm{min}) = h^{-}(x)$, $\psi(x,y_\mathrm{max}) = 
h^{+}(x)$.
These boundary conditions can be approximated as
\begin{equation}\label{bconds}
\sum_{m=0}^M \widetilde{\psi}_{nm} \delta^{\pm}_m = 
\widetilde{h}{}^{\pm}_n,
\qquad
\sum_{n=0}^N \widetilde{\psi}_{nm} \delta^{\pm}_n = 
\widetilde{f}{}^{\pm}_m,
\end{equation}
where one--dimensional Chebyshev projections appear on the right
hand sides. There are $2(N+1)+2(M+1)$ boundary conditions in
(\ref{bconds}), but we now show that they are not all linearly
independent due to double counting of corner conditions. The value
of $\psi(x_\mathrm{max},y_\mathrm{max}) = h^+(x_\mathrm{max})
= f^+(y_\mathrm{max})$ can be written as a linear
combination either of the $\widetilde{h}{}^{+}_n$ or of the
$\widetilde{f}{}^{+}_m$. This implies a homogeneous linear relationship
between the summations in the first and the second set of equations
in (\ref{bconds}). There are three other such linear relationships
that follow from the ways in which
$\psi(x_\mathrm{max},y_\mathrm{min})$,
$\psi(x_\mathrm{min},y_\mathrm{max})$,
and $\psi(x_\mathrm{min},y_\mathrm{min})$ can each be expressed
either in terms of sums of $\widetilde{h}{}^\pm_n$ or of
$\widetilde{f}{}^\pm_m$. Therefore, the number of independent
equations in (\ref{bconds}) is $2(N+1)+2(M+1)-4=2(N+M)$,
precisely equal to the number of zero rows.

All methods for eliminating the four linear
dependencies in (\ref{bconds}) should yield comparable
accuracy for a numerical solution. However, dropping the four
highest--mode equations ($n = N$ for $\pm$ in the left equation,
and $m=M$ for $\pm$ in the right equation) is inconsistent,
as confirmed by numerical experiments.
Indeed, by throwing out the highest mode on all four edges,
one loses information about the corner values. We have chosen
to reduce (\ref{bconds}) in a consistent way which is at the
same time convenient for our direct product representation of the
rectangular region. We use all $2N+2$ of the first set of
equations in (\ref{bconds}), i.e.~those equations involving
the $\widetilde{h}{}^{\pm}_n$. The Dirichlet vectors $\delta^\pm_m$
associated with these conditions are placed within the first two
rows of each block in the coefficient matrix (which does not
increase the bandwidth of the coefficient matrix beyond the
block--diagonal). We must then dispense with four equations
from the second set, two for $-$ (left) and two for $+$ (right),
and so drop those equations for which $m=M-1$ and $m=M$. This
corresponds to dropping the two highest modes for both left and
right Dirichlet conditions, although we have experimented with
dropping other modes and, as expected, found little difference.
From knowledge of all the $\widetilde{h}{}^\pm_n$ (information
determining all four physical corner values) and the
$\widetilde{f}{}^\pm_m$ for $m=0,\cdots,M-2$ we may recover the
four values for $\widetilde{f}{}^\pm_{M-1}$ and $\widetilde{f}{}^\pm_M$,
using the four equations for the corner values expressed as linear
combinations of the $\widetilde{f}{}^{\pm}_m$.
The Dirichlet vectors $\delta^\pm_n$ associated with the remaining
left/right boundary conditions are then placed in the remaining
$2M-2$ rows of the first two blocks (they reach across block columns
and so affect the bandwidth of the coefficient matrix quite a bit).
Values appearing on the righthand side of the above boundary conditions, 
all $\widetilde{h}{}^{\pm}_n$ and $\widetilde{f}{}^{\pm}_m$ except
for $\widetilde{f}{}^{\pm}_{M-1}$ and $\widetilde{f}{}^{\pm}_{M}$,
replace the appropriate zero entries of the source
$B^2_{x[2]}\otimes B^2_{y[2]}\widetilde{\mathbf{g}}$.

Although, our preconditioning of the rectangular operator has
resulted in a sparse matrix (while the original matrix corresponding
to the $D$s is dense), the issue of condition number is more
subtle. Clearly, the $B$s and $D$s are in some sense inverses of each
other. Notice that as an infinite dimensional matrix $B_{[1]}$ has the
action: $Q_0^{N-1} \rightarrow Q_1^{N}$, with $Q_k^p$ denoting the
vector subspace of spectral coefficients corresponding to Chebyshev
expansion from degree $k$ through degree $p$ polynomials. Suppose we
consider the $(N+1)$--by--$(N+1)$ truncation of $B_{[1]}$, and further
that we delete the first row and last column of this square matrix,
thereby obtaining an $N$--by--$N$ matrix $\bar{B}$. Likewise, we take
the $(N+1)$--by--$(N+1)$ truncation of $D$, and delete its first column
and last row to obtain $\bar{D}$. Then we may view $\bar{D}: Q_1^{N}
\rightarrow Q_0^{N-1}$. Taken as square matrices with the same domain
and range, $\bar{B}$ and $\bar{D}$ are nonsingular and inverses of
each other, whence have the same condition number.\footnote{Recall
that the condition number of a matrix $A$ is $\kappa(A) =
\|A^{-1}\|\|A\|$.} Such an argument can produce 
nonsingular matrices $\bar{B}^2$ and $\bar{D}^2$, also inverses of 
each other with the same condition number. Therefore, although
we are ignoring the critical issue of boundary conditions, 
passing from a coefficient matrix with symbolic form 
$I \otimes D^2 + D^2 \otimes I$ (corresponding to the Laplacian 
part of the operator) to one with the form $B^2 \otimes I + I 
\otimes B^2$ is not clearly advantageous insofar as the 
conditioning of the resulting linear system is concerned. 
Nevertheless, one might expect that a better distribution of 
eigenvalues for the form $B^2 \otimes I + I \otimes B^2$ would lead 
to faster convergence were we using an iterative solver such as 
GMRES\cite{Kelley}.

\subsection{Sparse formulation on annular domains}
In our 2d HRWE $L\psi=g$, 
we now take 
 $x = a + \rho\cos\theta$ and $y = b +
\rho\sin\theta$, where the ``hole'' is located at $(a,b)$,
which is either the center of annulus $H$ or the center of
$A$. In terms of 
\begin{equation}
F(\theta) = a\sin\theta-b\cos\theta,\qquad
G(\theta) = a\cos\theta
+b\sin\theta,
\end{equation}
the HRWE operator is
\begin{equation}
L =
\partial_\rho^2 +\rho^{-1}\partial_\rho 
+ \rho^{-2}\partial_\theta^2
-\Omega^2 \big[F(\theta)\partial_\rho 
+ (1+\rho^{-1}G(\theta))\partial_\theta
\big]^2\, .
\end{equation}
To prepare for the integration preconditioning, we use
$F'(\theta) = G(\theta)$ and $G'(\theta) = -F(\theta)$ along
with repeated appeals to the Leibniz rule in order to obtain
\begin{equation}
\rho^2 L\psi=\rho^2g
\end{equation}
with
\begin{eqnarray}
\rho^2 L & = & 
\partial^2_\rho\rho^2 (1-\Omega^2F^2) 
+\partial_\rho \rho[-3 + \Omega^2(2F^2
+ G^2 + \rho G)]
\nonumber \\
& &
+ \partial^2_\theta[1-\Omega^2(\rho+G)^2]
+\Omega^2\partial_\theta \rho F
-2\Omega^2\partial_\theta\partial_\rho \rho F(\rho+G)
\nonumber \\
& & +1-\Omega^2(G^2 +\rho G).
\end{eqnarray}

We now represent the solution on the annular subdomain in terms of
a truncated Fourier--Chebyshev expansion,
\begin{align}
\mathcal{P}_{M,N}\psi(\rho,\theta) = & 
\sum_{n=0}^N 
\widetilde{\psi}_{0n} T_n(\xi(\rho))
\nonumber \\
+ & \sum_{k=1}^{\frac{1}{2}M}\sum_{n=0}^N
\big[\widetilde{\psi}_{2k-1,n}\cos(k\theta)+
\widetilde{\psi}_{2k,n}\sin(k\theta)\big]T_n(\xi(\rho)),
\end{align}
where for simplicity we have here chosen $M$ even. For the direct 
product representation, 
\begin{equation}\label{annularordering}
\widetilde{\boldsymbol{\psi}}((N+1)m+n\big) = \widetilde{\psi}_{mn},
\end{equation}
which means each Fourier mode corresponds to its own
$(N+1)$--by--$(N+1)$ Chebyshev block.  We therefore have the following
matrix representation $\mathcal{L}$ of $\rho^2 L$:
\begin{eqnarray}
\mathcal{L} & = &
(I_\theta-\Omega^2\mathsf{F}^2)
\otimes D^2_\rho A_{\rho}^2
\nonumber \\
& & 
+\left(-3I_{\theta} +
2\Omega^2 \mathsf{F}^2
+ \Omega^2 \mathsf{G}^2
-2\Omega^2 D_\theta\mathsf{F}\mathsf{G}
\right) \otimes D_\rho A_{\rho}
\nonumber \\
& &
+ \Omega^2 \left(\mathsf{G}
-2 D_\theta\mathsf{F}
\right)
 \otimes
 D_\rho A_{\rho}^2
-\Omega^2 D^2_{\theta}\otimes A_{\rho}^2
\nonumber \\
& & 
+ \Omega^2 \left[D_\theta\mathsf{F}
- \left(I_\theta+ 2D^2_{\theta}\right)\mathsf{G}\right] \otimes
A_{\rho}
\nonumber \\
& &
+ \left(I_\theta+ D^2_{\theta}
\right)\left(I_\theta-\Omega^2\mathsf{G}^2\right) \otimes I_\rho\,.
\label{calL}
\end{eqnarray}
where san serif $\mathsf{F}$ and $\mathsf{G}$ denote 
matrices in the Fourier sin/cos basis corresponding to
multiplication by $F$ and $G$. The entries of these matrices
are determined by standard trigonometric addition--of--angle 
formulas, such as $2\sin\alpha\cos\beta = 
\sin(\alpha + \beta) + \sin(\alpha - \beta)$. For the scenario of 
the 3d HRWE, such a spectral approximation for a spherical shell 
around an inner hole would be rather more problematic (even if the 
analogous matrix is not explicitly formed). Indeed, for that 
scenario we would need to contend with Wigner--Clebsch--Gordon
coefficients arising from products of spherical harmonics $Y_{\ell m}$. 

Since the spectral differentiation matrices $D_\rho$ and $D_\rho^2$
are dense upper triangular, passage to a sparse--matrix formulation 
of the problem for an annulus necessarily requires that we 
apply the integration matrix $B^2_{\rho[2]}$. By contrast, since the 
matrices $D_\theta$ and $D^2_\theta$ are already banded or diagonal, 
we will achieve a sparse formulation whether or not we precondition 
with $\theta$ integration. {\em In fact we have employed $\theta$ 
preconditioning}, since it might well yield a better distribution
of eigenvalues and better equilibration properties. Nevertheless, 
for simplicity will here ignore IPC in $\theta$, which, in any case, 
is logically different from IPC in $\rho$ in that no boundary 
condition is associated with the periodic direction.
The preconditioning integration matrix is then $\mathcal{B} = 
I_{\theta}\otimes B^2_{\rho[2]}$, and its application onto 
(\ref{calL}) yields
\begin{eqnarray}
\mathcal{B}\mathcal{L} & = &
(I_\theta-\Omega^2\mathsf{F}^2)
\otimes I_{\rho[2]} A_{\rho}^2
\nonumber \\
& &
+\left(-3I_{\theta} +
2\Omega^2 \mathsf{F}^2
+ \Omega^2 \mathsf{G}^2
-2\Omega^2 D_\theta\mathsf{F}\mathsf{G}
\right) \otimes B_{\rho[2]} A_{\rho}
\nonumber \\
& &
+ \Omega^2 \left(\mathsf{G}
-2 D_\theta\mathsf{F}
\right)
 \otimes
 B_{\rho[2]} A_{\rho}^2
-\Omega^2 D^2_{\theta}\otimes B^2_{\rho[2]}A_{\rho}^2
\nonumber \\
& &
+ \Omega^2 \left[D_\theta\mathsf{F}
- \left(I_\theta+ 2D^2_{\theta}\right)\mathsf{G}\right] \otimes
B^2_{\rho[2]}A_{\rho}
\nonumber \\
& &
+ \left(I_\theta+ D^2_{\theta}
\right)\left(I_\theta-\Omega^2\mathsf{G}^2\right) \otimes B^2_{\rho[2]}\,.
\end{eqnarray}
The right hand side
of $\rho^2L\psi=\rho^2g$ is now represented by $\mathcal{B}\cdot(I_\theta
\otimes A_{\rho}^2) \widetilde{\mathbf{g}} = (I_\theta \otimes
B^2_{\rho[2]} A_{\rho}^2)\widetilde{\mathbf{g}}$.  For the outer
annulus we have $(a,b) = (0,0)$, and so $\rho = r$ and $\theta =
\varphi$. Also for this case $F(\theta) = 0 = G(\theta)$, and
$\mathcal{L}$ in (\ref{calL}) reduces to
\begin{equation}
\mathcal{L} = 
I_\varphi\otimes D^2_r A^2_r
-3 I_\varphi\otimes D_r A_r
+  D^2_\varphi \otimes (I_r-\Omega^2A^2_r) + I_\varphi\otimes I_r.
\end{equation}
We therefore find
\begin{equation}
\mathcal{B}\mathcal{L} =
I_{\varphi}
\otimes I_{r[2]}A_r^2
-3I_{\varphi}
\otimes B_{r[2]}A_r
+ D^2_{\varphi} \otimes 
B^2_{r[2]}
\left(I_r - 
\Omega^2A_r^2
\right)
+ I_{\varphi}
\otimes B^2_{r[2]}
\label{BLinA}
\end{equation}
as the preconditioned matrix for the outer annulus.

Turning to the issue of boundary conditions, we first note that for 
$\mathcal{B}\mathcal{L}$ the first two rows of each block have 
all zero entries. Into these rows we therefore place the 
$\tau$--conditions,
\begin{equation}
\sum_{n=0}^N \widetilde{\boldsymbol{\psi}}_{mn}\delta_n^{\pm}
= \hat{h}{}^\pm_m
\end{equation}
(and similarly for Neumann conditions).
Here $\hat{h}{}^\pm_m$ is the Fourier transform of the
boundary conditions $h^\pm(\theta)$, for example with 
$h^-(\theta) = \psi(\rho_\mathrm{min},\theta)$. Notice that
the correct number of zero entries in the preconditioned 
source correspond to the inhomogeneity $\hat{h}{}^\pm_m$.
For the radiation ``$p,q$ boundary conditions'' (\ref{rmaxuBCs}) we 
are only concerned with the outer annulus and the simpler operator 
(\ref{BLinA}). In this case, the overall 
 $(N+1)(M+1)$--by--$(N+1)(M+1)$ matrix $\mathcal{BL}$ is block 
diagonal, with $(M+1)$ blocks. Each block has the structure
\begin{equation}
\left[
\begin{array}{c}
\mathbf{0}\\
\mathbf{0}\\
\hline
\mathcal{BL}^k
\end{array}
\right]
\end{equation}
for some Fourier wave number $k$. Here the $\mathbf{0}$ represents a
row of zeros, and $\mathcal{BL}^k$ is a nonzero $(N-1)$--by--$(N+1)$ 
submatrix. As the radiation boundary conditions couple sine and 
cosine modes of the same wave number, we must consider the two 
consecutive blocks (one cosine and the other sine) associated with Fourier 
wave number $k \neq 0$. As depicted in the last equation, each of the two 
blocks has all zero entries in its first two rows. The $p,q$ boundary 
conditions are enforced by choosing the overall block 
neighborhood as follows,
\begin{equation}
\left[
\begin{array}{c|c}
\begin{array}{c}
\delta^-\\
p\delta^+\\
\mathcal{BL}^k
\end{array}
&
\begin{array}{c}
\mathbf{0}\\
R\nu^+ + q\delta^+\\
\mathbf{0}
\end{array}
\\
\hline
\begin{array}{c}
\mathbf{0}\\
R\nu^+ + q\delta^+\\
\mathbf{0}
\end{array}
&
\begin{array}{c}
\delta^-\\
-p\delta^+\\
\mathcal{BL}^k
\end{array}
\end{array}
\right],
\label{Ablock}
\end{equation}
where $\mathbf{0}$ represents either a row or a $(N-1)$--by--$(N+1)$ submatrix 
of zeros. Boundary conditions for $k=0$ (the zero mode) are easier to enforce 
(only a single block need be considered) and are handled similarly.

\subsection{Gluing of subdomains}
So far we have described individual rectangular and annular subdomains
(and their associated $\tau$--conditions) as if these subdomains were 
decoupled. In fact, we ``glue together'' all or most of the subdomains 
shown in Fig.~\ref{domaindecomp}. This gluing takes two forms: (i) imposing 
matching conditions for adjacent rectangles and (ii) imposing matching
conditions for the overlap between an annulus and a set of rectangles.
Before describing each case in more detail, we comment on how such 
gluing is reflected in the overall linear system. Let 
$\widetilde{\boldsymbol{\psi}}{}^H$ and 
$\widetilde{\boldsymbol{\psi}}{}^A$ represent the vectors
of Fourier--Chebyshev expansion coefficients associated with the 
spectral representation of the solution on the annuli $H$ and $A$.
Similarly, let $\widetilde{\boldsymbol{\psi}}{}^j$ represent the vector of 
double--Chebyshev expansion coefficients associated with the
spectral representation of the solution on the $j$th rectangle
(with $1 \leq j \leq 8$). The overall set of unknowns is then the
concatenation $\widetilde{\mathbf{\Psi}}=
(\widetilde{\boldsymbol{\psi}}{}^H,
\widetilde{\boldsymbol{\psi}}{}^1,
\widetilde{\boldsymbol{\psi}}{}^2,
\widetilde{\boldsymbol{\psi}}{}^3,
\widetilde{\boldsymbol{\psi}}{}^4,
\widetilde{\boldsymbol{\psi}}{}^5,
\widetilde{\boldsymbol{\psi}}{}^6,
\widetilde{\boldsymbol{\psi}}{}^7,
\widetilde{\boldsymbol{\psi}}{}^8, 
\widetilde{\boldsymbol{\psi}}{}^A)^t$, which satisfies (for the linear 
problem) the spectral matrix form of Eq.~(\ref{introHRWE})
\begin{equation}
\mathcal{M}\widetilde{\mathbf{\Psi}} = 
\mathcal{B}\widetilde{\mathcal{G}},
\end{equation}
where $\widetilde{\mathcal{G}}$ is a similar concatenation of the sources 
$\widetilde{\mathbf{g}}$ on 
the individual subdomains and the $\mathcal{B}$ indicates integration
preconditioning on all subdomains. Symbolically, the coefficient
matrix $\mathcal{M}$ is $\mathcal{B}\mathcal{L}$, here with $\mathcal{L}$
standing for the spectral representation of the HRWE operator $L$ on the
whole two center domain.

Each of the ten subdomains (annuli $A$ and $H$, as well as rectangles
1--8) in Fig.~\ref{domaindecomp} are represented by one of ten
super--blocks ($H$--$H$, 1--1, 2--2, $\cdots$,
8--8, $A$--$A$) which sit along the diagonal of the overall super
matrix $\mathcal{M}$ representing the PDE on the whole two center
domain. We use the term ``super--block'' here since the matrix
corresponding to each subdomain arises, as we have seen, from a direct
product structure (and so could be viewed as already in a block
form). The supplementary equations needed for gluing are placed within
existing zero rows in the same manner as with the
$\tau$--conditions. However, the gluing conditions stretch beyond the
super--block diagonal, since they are linear relationships between the
spectral expansion coefficients on two (or more) separate subdomains. For
example, the gluing together of subdomains 1 and 2 (which share a common
edge) involves not only filling rows within the 1--1 and 2--2
super--blocks along the diagonal of $\mathcal{M}$, but also filling
rows within the 1--2 and 2--1 off--diagonal super--blocks.

\subsubsection{Gluing of rectangles to rectangles}\label{subsubsec:glueRecs}
For rectangles which meet at an edge we require both continuity in
$\psi$ and its first derivative $\partial\psi/\partial \nu$ (normal to
the matching edge). We impose these requirements strongly, that is to
say at the level of the numerical solution itself. Consider, for
example, rectangles 1 and 2 in Fig.~\ref{domaindecomp}, which as
indicated share the edge $y = \rho_\mathrm{min}$, where
$\rho_\mathrm{min}$ is the radius of the inner hole (the depicted
excised region). We require that values of $\psi(x,\rho_\mathrm{min})$
and $\partial\psi/\partial y(x,\rho_\mathrm{min})$ agree at the
Chebyshev--Gauss--Lobatto collocation points $x(\xi_i)$ whether these
values are computed using the spectral coefficients of rectangle 1 or
those of rectangle 2. For simplicity, we assume that the number $N_1$ of 
spectral elements for rectangle 1 is the same as that $N_2$ for 
rectangle 2, so that the matching equations are simply
\begin{equation}
\sum_{m = 0}^{M_1} \widetilde{\psi}{}^1_{nm} \delta_m^+ =
\sum_{m = 0}^{M_2} \widetilde{\psi}{}^2_{nm} \delta_m^-,
\qquad
 \sum_{m = 0}^{M_1} \widetilde{\psi}{}^1_{nm} \alpha_1\nu_m^+ =
-\sum_{m = 0}^{M_2} \widetilde{\psi}{}^2_{nm} \alpha_2 \nu_m^-,
\label{1-2tau}
\end{equation}
for each value of $n$. 
The $\alpha$ factors here are the scalings
of the Neumann vectors
 that are necessary 
since the range of $y$ may not be $[-1,1]$. (See the discussion 
following 
Eq.~(\ref{numinus}).)
These matching conditions are reflected in the overall 
matrix $\mathcal{M}$ as follows. As the super--block corresponding to 
each of the subdomains 1 and 2 has been preconditioned in the described 
fashion, each has a collection of zero rows in which we place the 
matching conditions.  In, say, the zero rows belonging to the super--block 
1--1, we insert the first set of conditions given in (\ref{1-2tau}). In 
the zero rows belonging to the super--block 2-2, we similarly place the 
Neumann conditions, the second set of conditions given in (\ref{1-2tau}). 
We note that this filling of zero rows to achieve the required
matching does not affect the inhomogeneity, as these are homogeneous 
conditions (a linear sum of expansion coefficients for one subdomain plus 
a linear sum of expansion coefficients for another is set equal to zero). 

In practice, the gluing procedure is plagued by the same sort of
issues that arose in connection with Eq.~(\ref{bconds}). Ultimately
due to redundant counting at corners, the full set of matching
conditions responsible for the gluing of all subdomains (including annuli)
are not linearly independent. Thus we find that the number of apparent
matching (and boundary) conditions is greater than the number of zero rows
available in the super matrix $\mathcal{M}$ to hold the such conditions.
As we did for a single rectangle, we have chosen to consistently eliminate
linear dependencies at left--right edges/interfaces, as illustrated in the
next paragraph. The number of zero rows in $\mathcal{M}$ always equal to the
number of linearly independent matching (and boundary) conditions, and
in principle any two ways of implementing them should be equivalent.
While the choice of implementation does of course affect the structure of
the super matrix $\mathcal{M}$, it should not greatly affect the
accuracy of our numerical solutions.

Let us turn to our illustrative example, the matching, along the
vertical edge $x = x_H + \rho_\mathrm{min}$, of rectangles 2 and 3 in
Fig.~\ref{domaindecomp}. In the matrix for rectangle 2
(super--block 2--2 of the overall super matrix ${\mathcal M}$)
we have reserved $2M_2-2$ rows in the first two blocks for enforcing
boundary (or matching) conditions at the left and right edges. We will
now have $M_2 - 1$ of these available for matching at the left edge of
rectangle 2. Likewise, in the matrix for rectangle 3 (super--block 3--3
of the overall super matrix ${\mathcal M}$) we will have
$M_3 - 1$ available rows for the matching along the right edge of
rectangle 3. We take $M_1 = M_2 = M$, so that we have a total of
$2M-2$ rows in the overall super matrix ${\mathcal M}$ available to
explicitly enforce the matching of rectangles 2 and 3. The full set of
equations for gluing along this edge are analogous to (\ref{1-2tau}), but with
$N$s and $M$s interchanged,
\begin{equation}
\sum_{n = 0}^{N_2} \widetilde{\psi}{}^2_{nm} \delta_n^- =
\sum_{n = 0}^{N_3} \widetilde{\psi}{}^3_{nm} \delta_n^+,
\qquad
 \sum_{n = 0}^{N_2} \widetilde{\psi}{}^2_{nm} \alpha_2\nu_n^- =
-\sum_{n = 0}^{N_3} \widetilde{\psi}{}^3_{nm} \alpha_3 \nu_n^+.
\label{2-3tau}
\end{equation}
We follow a protocol similar to the convenient one outlined for fixing
Dirichlet boundary conditions for a single isolated rectangle. Whereas
we have chosen to devote a zero row in $\mathcal{M}$ to each of the
$2N+2$ matching conditions (\ref{1-2tau}) for a top--bottom gluing, for a left--right
gluing we devote only $M-1$ zero rows for each set in (\ref{2-3tau}),
that is $2M-2$ rows in all. This means dropping the last
two (highest mode) equations from each set.
Since we work with the transform of the matching conditions, dropping the
highest two modes does not mean we are not enforcing matching at a
particular point along the interface. Moreover, although we have chosen
not to reflect all of the explicit matching conditions (\ref{2-3tau})
in $\mathcal{M}$, extra matching of rectangles 2 and 3 is afforded by
other matching conditions in the overall linear system. For example,
along both the top edge of rectangle 2 and top edge of rectangle 3, we
enforce a full set of boundary (or matching) conditions with Dirichlet
data coming from annulus $A$. Since that data will be smooth along the
full length of the combined top edge for both rectangles 2 and 3,
these top--edge boundary conditions effectively yield extra matching
for the 2--3 vertical gluing, in particular enforcing both continuity
and differentiablity at the physical corner point common to each
rectangle.

Another complication in gluing is the implementation of symmetry.
Notice that the left edge of rectangle 4, for example, must be {\em flipped}
and then glued to the left edge of rectangle 7. Similar flipped gluing must also
be considered for subdomains 5 and 6. We enforce such flipped gluing of edges
by demanding continuity in $\psi$ only (since region 7 is already matched to
region 6), and filling the available zeros rows of the 4--4 super--block of
$\mathcal{M}$ accordingly (as in this case the 6--6 and 7--7 super--blocks
would no longer have appropriate free rows available).

Let us provide a brief sketch of rectangle gluing without the
``conforming assumption.'' Were we gluing, say, rectangle 1
to rectangle 2 (a top--bottom glue) with $N_1 \neq N_2$, we
would deal with more complicated equations than (\ref{1-2tau}). The
second equation in (\ref{1-2tau}), for example, could be written as
\begin{equation}
\sum_{m=0}^{M_1} \widetilde{\psi}{}^1_{nm} \alpha_1 \nu^+_m
= \widetilde{h}{}^+_n,
\label{new12eqns}
\end{equation}
where $\widetilde{h}{}^+_n$ is simply the explicit sum on
the righthand side when $N_1 = N_2$. However, were
$N_1 \neq N_2$, it would then arise as the $x$--Chebyshev
transform of the vector
\begin{equation}
h^+_n = \sum_{p,q = 0}^{N_2,M_2}
\widetilde{\psi}{}^2_{pq}
T_p(\xi^2(x_n))\left.\left[ \frac{\mathrm{d}}{\mathrm{d} y}
T_q(\eta^2(y))\right]\right|_{y=\rho_{\mathrm{min}}}.
\end{equation}
Here the $x_n$ are the $N_1+1$ (scaled) Chebyshev--Gauss--Lobatto
points for rectangle 1, and $(\xi^2(x),\eta^2(y))$ are the
linear functions mapping rectangle 2 to the square $[-1,1]
\times [-1,1]$. As there would be $N_1 + 1$ of the equations
(\ref{new12eqns}), they would be placed in zero rows
corresponding to the 1--1 super--block of $\mathcal{M}$, and
would stretch across the 1--1 and 1--2 super--blocks. Equations for
continuity similar to the first set of equations in (\ref{1-2tau}), 
but using
the $x$--Chebyshev transform for rectangle 2, would determine
entries in $\mathcal{M}$ stretching across the 2--1 and 2--2 
super--blocks. The roles of 1 and 2 could be interchanged. The 
situation is nearly the same for a left--right
gluing, modulo the issue of dropping the two highest modes.

The rows in the overall super--matrix $\mathcal{M}$ responsible for all of
the rectangle--to--rectangle gluing result in $\mathcal{M}$ having a large
condition number $\kappa(\mathcal{M})$. Indeed, for the truncations we later 
consider in our numerical experiments, the matrix for the elliptic region 
(which includes the $H$--$H$ super--block as well) has a reciprocal condition number 
{\tt RCOND} (a diagnostic determined by {\tt dgesvx}) in the $10^{-6}$ to 
$10^{-10}$ range.
%
%
%
As we will see in Sec.~\ref{subsec:2dsims}, such a large condition number 
leads to an accuracy problem for higher truncations. 
In an attempt to mitigate this problem, we have experimented with other
strategies for rectangle gluing, for example, imposing point--space
rather spectral versions of the matching and boundary conditions.
Another possibility, suggested by a referee, and for which we had
earlier carried out some numerical experiments, is to allow
for overlap between adjacent rectangles. In this case the gluing procedure
is similar to the described rectangle--annulus gluing, with the key advantage
that all Neumann vectors can be done away with (this could lead to improved
conditioning). The disadvantage is that significant overlap is typically
required for good conditioning, and, therefore, more of the physical space
is doubly covered. Beyond an increase in resources, such double cover also
further complicates the gluing of the overall concatenation of rectangles
to the annuli. Nevertheless, this idea deserves further study. To sum up,
we have not exhaustively tried every approach, and our current one is
quite possibly not optimal, although we believe it will be hard to do
significantly better. Nevertheless, by
switching from {\tt dgesv} to {\tt dgesvx}, we are able to achieve the 
requisite accuracy for our experiments. The routine {\tt dgesvx} includes 
equilibration (which changes {\tt RCOND} to the $10^{-4}$ to $10^{-6}$ 
range) and iterative refinement\cite{Demmel}. Moreover, with our
approach we have carried out the experiment (``$L=5$ with linear 
mappings") documented in Sec.~4.1 of \cite{PKSTelliptic}, the Laplace 
equation for the exact solution $\ln(x^2+y^2)$ on a $10 \times 10$ 
square with an excised $1 \times 1$ inner square ``hole.''
With {\tt dgesv} we attain a best error measure which is about three 
orders of magnitude larger than the corresponding one reported in 
Fig.~3 of that reference. However, with {\tt dgesvx} we find
nearly identical results, and even a slightly better best error measure.

\subsubsection{Gluing of annuli to rectangles} The annuli $H$ and $A$
depicted in Fig.~\ref{domaindecomp} overlap multiple rectangles, and
for this overlap the issue of gluing is complicated. Since the issue 
is essentially the same for the gluing of $H$ to rectangles 1--8 or $A$ 
to rectangles 1--4 and 6--8, let us here focus on the first case. For 
example, part of the outer edge of annulus $H$ sits in rectangle 1; we 
require that the values of $\psi$ along this outer edge of $H$ agree 
whether they are
computed with the spectral representation of $\psi$ on $H$ or with
the spectral representation of $\psi$ on rectangle $1$. Similarly,
the values of $\psi$ at the left edge of rectangle 1 must be the
same whether computed using the rectangle 1 or $H$ spectral 
representation. That is to say, along these boundaries we 
demand agreement in the numerical solution whether determined by the 
expansion coefficients $\widetilde{\boldsymbol{\psi}}{}^H$ of the 
annulus or the coefficients $\widetilde{\boldsymbol{\psi}}{}^1$ for 
rectangle 1. An essential difference should be noted between 
this kind of gluing, and the gluing of two rectangles at an edge:
Here we use Dirichlet matching along two curves rather than 
Dirichlet plus Neumann matching along a single shared boundary
curve.

For the outer circular boundary $\rho = \rho_\mathrm{max}$ of the
annulus $H$, we enforce matching at each of a set of Fourier
collocation points
$\{\theta_m : m = 1,2,\cdots, M_H\}$ through the following set of
equations:
\begin{equation}\label{outedge1}
\sum_{n=0}^{N_H} \widetilde{\psi}{}^H_{nm}\delta^+_n = \hat{h}^+_m,
\end{equation}
where now the $\hat{h}^+_m$ are not fixed Dirichlet values. They now 
arise via Fourier transform of
\begin{equation}\label{outedge2}
h^+_m = \sum_{p,q=0}^{N_j,M_j} \widetilde{\psi}{}^j_{pq} 
T_p(\xi^j(x_m)) 
T_q(\eta^j(y_m)).
\end{equation}
Here the $j$th rectangle contains the point $(x_m,y_m)$ corresponding to
the collocation point $(\rho_\mathrm{max},\theta_m)$. These equations are 
placed within the zero rows of the $H$--$H$ super--block. The explicit
matching equations can be captured by expressing the Fourier transform
$\hat{h}^+_m = \sum_{k=1}^{M_H}\mathcal{F}_{mk} h^+_k$ as a matrix relation
in terms of the $\widetilde{\psi}{}^j_{pq}$.

The 
matching along one of the inner edges of a rectangle follows a similar
pattern.
As a concrete example, take again the inner edge of rectangle 1, and the
condition
\begin{equation}
\sum_{n=0}^{N_1} \widetilde{\psi}{}^1_{nm}\delta^-_n = \widetilde{f}_{m},
\end{equation}
where the $\widetilde{f}_{m}$ are now not fixed Dirichlet values. Rather,
they arise as the $y$--Chebyshev transform of the vector
\begin{align}
f_m & = \sum_{p=0}^{N_H}
\widetilde{\psi}{}^H_{0p} T_p(\xi(\rho_m))
\nonumber \\
& + \sum_{k=1}^{\frac{1}{2}M_H}\sum_{p=0}^{N_H}
\big[\widetilde{\psi}{}^H_{2k-1,p}\cos(k\theta_m)+
\widetilde{\psi}{}^H_{2k,p}\sin(k\theta_m)\big]T_p(\xi(\rho_m)).
\end{align}
Here the point $(\rho_m,\theta_m)$ corresponds to a Chebyshev--Gauss--Lobatto
collocation point $(x_\mathrm{min},y(\eta_m))$ along the inner edge, and
we have chosen $M_H$ as an even integer for simplicity. (Note that 
the  values 
of $\theta_m$ here, with $0\leq m\leq M_1$ have no direct relationship
to the Fourier collocation points $\theta_m$ used in Eqs.~(\ref{outedge1})
and (\ref{outedge2}).) Again,
via a matrix representation $\widetilde{f}_m = \sum_\ell \mathcal{C}_{m\ell}
f_\ell$ of the transform, we may express this matching condition more
directly. In any case, these conditions are placed within the available
zero rows of the 1--1 super--block of $\mathcal{M}$.

\subsection{Nonlinear model}\label{nonlinmod}

We turn to the 2d nonlinear HRWE discussed in \cite{WheKrivPri}, that is
\begin{equation}
L \psi + \eta h(\psi) = g,\qquad 
h(\psi) = \frac{\psi^5}{\psi^4 + \psi_0^4}.
\label{nonlin2dHRWE}
\end{equation}
Certain aspects of our treatment of the nonlinear term will likely not generalize 
to 3d. Indeed, we have retained a pure spectral method (coefficients as unknowns), 
whereas most modern approaches\cite{Kelley} towards solving nonlinear equations 
via Newton--Krylov spectral methods have centered around the construction of 
preconditioners within the context of pseudospectral methods (point values as 
unknowns). Nevertheless, as discussed in the concluding section, we expect 
that some aspects of our approach will carry over to 3d insofar as an outer 
spherical shell is concerned.

Again, we let $\widetilde{\mathbf{\Psi}}$ represent the vector of unknowns, 
that is the overall concatenation of 
the spectral coefficients on all subdomains. Using the inverse spectral
transformation available on each subdomain, we may produce from 
$\widetilde{\mathbf{\Psi}}$ a collection of point--space values $
\mathbf{\Psi} = \mathcal{F}^{-1}\widetilde{\mathbf{\Psi}}$ defined on 
each subdomain's spectral grid. Here we
are using $\mathcal{F}^{-1}$ to formally denote the process of inverse 
transformation on all subdomains. Our problem now is to solve
\begin{equation}
W[\widetilde{\mathbf{\Psi}}] = \mathcal{M}\widetilde{\mathbf{\Psi}} + 
\eta \mathcal{B}\mathcal{F} h\big(\mathcal{F}^{-1}\widetilde{\mathbf{\Psi}}\big) 
- \mathcal{B}\widetilde{\mathcal{G}} = 0,
\end{equation}
here with $\mathcal{B}$ representing the application of integration 
preconditioning on all subdomains, $\widetilde{\mathcal{G}}$ the 
concatenation of all source coefficients $\widetilde{\mathbf{g}}$, and 
$\mathcal{M} = \mathcal{B}\mathcal{L}$ the coefficient matrix for the
linear HRWE on the glued--together two center domain. The term $\mathcal{F} 
h\big(\mathcal{F}^{-1}\widetilde{\mathbf{\Psi}}\big)
= \mathcal{F}h\big(\mathbf{\Psi}\big)$ is the
spectral representation of the nonlinearity, where 
$h\big(\mathbf{\Psi}\big)$ is constructed pointwise on each 
subdomain's spectral grid.

Solving this nonlinear system of equations requires a Newton-Raphson
iteration, and we briefly describe how this has been implemented.
First, we note that by using both (\ref{TxTequalsT}) and
addition of angle formulas, such as $2\sin\alpha\cos\beta =
\sin(\alpha+\beta) + \sin(\alpha-\beta)$, we may convert the spectral
coefficients $\widetilde{\mathbf{u}}$ representing a function $u$ on a
particular subdomain into a matrix $C_u$ whose action on the
subdomain's vector space of spectral coefficients corresponds to
multiplication by $u$ in physical space. For example, on a rectangular 
domain, the resulting matrix would arise as
\begin{equation}
\sum_{n=0}^N\sum_{m=0}^M \widetilde{u}_{nm} 
T_n(\xi(x)) T_m(\eta(y)) \rightarrow 
C_u = \sum_{n=0}^N\sum_{m=0}^M \widetilde{u}_{nm}
A_{T_n} \otimes A_{T_m},
\end{equation}
where $A_{T_n} = T_n(A_x)$ is $(N+1)$--by--$(N+1)$, 
$A_{T_m} = T_m(A_y)$ is $(M+1)$--by--$(M+1)$, and each 
matrix has been appropriately scaled (as discussed at the
end of Section \ref{basicChebyForms}).
For both annuli and rectangles we have written
subroutines for fast computation of such matrices from input
coefficients. Then, given a concatenation $\widetilde{\mathbf{U}} =
(\widetilde{\mathbf{u}}{}^H, \widetilde{\mathbf{u}}{}^1,
\widetilde{\mathbf{u}}{}^2, \widetilde{\mathbf{u}}{}^3,
\widetilde{\mathbf{u}}{}^4, \widetilde{\mathbf{u}}{}^5,
\widetilde{\mathbf{u}}{}^6, \widetilde{\mathbf{u}}{}^7,
\widetilde{\mathbf{u}}{}^8, \widetilde{\mathbf{u}}{}^A)^t$ of the
spectral coefficients on all subdomains, we may also form a
(super--block diagonal) matrix $\mathcal{C}_u$ whose action on 
the vector space of spectral coefficients for the entire two center 
domain corresponds to multiplication by $u$ in physical space
over the entire two center domain.

Let $\widetilde{\boldsymbol{\Psi}}{}^\mathrm{old}$
represent the current Newton iterate. Then computing the next iterate
$\widetilde{\boldsymbol{\Psi}}{}^\mathrm{new} 
= \widetilde{\boldsymbol{\Psi}}{}^\mathrm{old} - \delta
\widetilde{\boldsymbol{\Psi}}{}$ requires that we solve the linear equation
\begin{equation}
\left\{
\mathcal{M} +
\eta\mathcal{B}\mathcal{C}_{h'(\boldsymbol{\Psi}^\mathrm{old})}
\right\}\delta\widetilde{\boldsymbol{\Psi}}{} = 
W[\widetilde{\boldsymbol{\Psi}}{}^\mathrm{old}],
\label{NewtonSolve}
\end{equation}
where the derivative function is
\begin{equation}
h'(\psi) = \frac{5\psi^4 \psi_0^4 + \psi^8}{\big[\psi^4 + \psi_0^4\big]^2}.
\end{equation}
The matrix $\mathcal{C}_{h'(\boldsymbol{\Psi}^\mathrm{old})}$
is constructed as follows: (i) obtain the point values 
$\mathbf{\Psi}^\mathrm{old} = 
\mathcal{F}^{-1}\widetilde{\mathbf{\Psi}}^\mathrm{old}$,
(ii) construct the set of point values $h'(\mathbf{\Psi}^\mathrm{old})$,
(iii) construct the concatenation of spectral coefficients
$\widetilde{\mathbf{U}} = \mathcal{F} h'(\mathbf{\Psi}^\mathrm{old})$,
(iv) using the coefficients $\widetilde{\mathbf{U}}$, form the matrix
$\mathcal{C}_{h'(\boldsymbol{\Psi}^\mathrm{old})}$ as outlined above.
We have taken advantage of the sparse nature of the preconditioner 
$\mathcal{B}$ to achieve fast matrix--matrix and matrix--vector multiplies, 
with the former required for quick computation of  the matrix--matrix
product $\mathcal{B}\mathcal{C}_{h'(\boldsymbol{\Psi}^\mathrm{old})}$
that is needed for each Newton iterate. Each such linear solve has been
performed directly with {\tt dgesv}.

We have found that the described implementation of Newton--Raphson iteration 
works well for the 2d model, and for a wide range of parameter 
choices we have not needed to include line searches. Nevertheless, we 
recognize that such an implementation is not likely a viable option for 3d. 
Indeed, for spherical shells the scalar spherical harmonics
$Y_{\ell m}$ are required for the angular basis functions. Beyond the
memory burden of forming matrices for the 3d problem, even
representation of multiplication by a nonlinear function in the
spectral $Y_{\ell m}$ basis is not as practical, because it would 
require extensive use of Wigner--Clebsch--Gordan coefficients (the simple
multiplicative properties embodied in addition--of--angle formulas are
no longer at one's disposal). However, in a Krylov approach one might use 
the spherical harmonic transform and inverse transform to the same effect, 
carrying out all multiplications in point space.

\section{Numerical analysis and results}\label{numerical}

This section considers the conditioning of our numerical problem on
the outer annulus $A$. It also presents our results in solving the
HRWE on various subdomains as well as on the whole two center domain.
In the first subsection we consider only a single Fourier mode 
in the outer annulus, and we study the condition number of the matrix 
that must be inverted. This analysis also pertains
to the analogous outer shell problem associated with solving the 3d
HRWE.  Indeed, as we have shown in the second section, the ODE and
boundary conditions governing the 2d and 3d mode equations are
essentially the same. The second subsection describes methodology and 
collects error tables associated with our numerical experiments in 
solving the linear and nonlinear 2d HRWE. A third subsection presents
preliminary results in solving the 3d HRWE.

\subsection{Condition numbers}
Consider the radial operator which stems from the linear helical
operator via a Fourier--series transform; consider also the radial
operator that arises in this way from the Laplace--Poisson
operator. With $k$ taken as the wave number, these radial operators
are respectively (see left hand side of Eq.~(\ref{helicalft}))
\begin{equation}
L^k = r^2 \frac{\mathrm{d}^2}{\mathrm{d} r^2}
+ r \frac{\mathrm{d}}{\mathrm{d} r}
- k^2(1-\Omega^2 r^2), \qquad \mathrm{(Helical\; operator)},
\end{equation}
with Dirichlet boundary conditions at the inner radius and
exact outgoing radiation boundary conditions at the outer
radius (D--R conditions), and
\begin{equation}
P^k = r^2 \frac{\mathrm{d}^2}{\mathrm{d} r^2}
+ r \frac{\mathrm{d}}{\mathrm{d} r}
- k^2, \qquad \mathrm{(Poisson\; operator,}\;\Omega = 0).
\end{equation}
For the Poisson operator we examine both
Dirichlet--Dirichlet (D--D) and Dirichlet--Neumann (D--N) boundary
conditions. As the Laplacian is translation invariant, the radial 
Poisson--Laplace operator $P^k$ can be put into the same Fourier form 
wherever the annulus lies. The HRWE mixes Fourier modes if the 
center of the angular coordinates is offset from the center of rotation, 
as in the case of the hole subdomain $H$. 

Here, however, we will work with the outer 
annulus $A$ for which the HRWE itself does not mix modes, 
although, as we have seen, the radiation boundary condition does mix 
cos and sin modes of the same wave number. Therefore, we essentially 
have an ODE setting in which the conditioning properties of IPC are 
documented\cite{CHHT}. We point out that Ref.~\cite{CHHT} also
refers to the conditioning of a ``helical operator.'' However,
the helical operator of Ref.~\cite{CHHT} arises in the context of 
the Navier--Stokes equations and is of course different from the 
one we consider.

Before numerically probing the issue of conditioning for the 
operators above, let us present a heuristic argument as to why IPC 
achieves improved conditioning in certain ODE settings, including
as special cases those of immediate interest to us. Suppose we 
have the equation
\begin{equation}
\frac{d^2}{dr^2} \xi + \frac{d}{dr} s_1(r)\xi + s_0(r)\xi = g,
\label{generalODE}
\end{equation}
where the $s_1(r)$ and $s_0(r)$ are polynomials
and the equation must be supplemented with appropriate 
boundary conditions. Here, one should think of $\xi(r)$ as a mode,
either $\hat{\psi}_k(r)$ or $\hat{\psi}_{\ell m}(r)$. Performing 
the IPC 
technique on the ODE (\ref{generalODE}), we pass from a spectral 
approximation of the equation based on differentiation matrices,
\begin{equation}
\big[D^2 + D S_1(A) + S_0 (A)\big]\boldsymbol{\xi} = \mathbf{g},
\label{diffrep}
\end{equation}
to the following matrix representation:
\begin{equation}
\left[I_{[2]} + B_{[2]} S_1(A)  + B^2_{[2]} S_0(A)\right]
\boldsymbol{\xi}
= B^2_{[2]}\mathbf{g},
\label{intrep}
\end{equation}
with two available rows of zero in which to put $\tau$--conditions
specifying the boundary conditions to be applied. In either
representation, $S_1(A)$ and $S_0(A)$ are the matrices which 
correspond to multiplication by $s_1(r)$ and $s_0(r)$ in the
Chebyshev basis. In the passage
from (\ref{diffrep}) to (\ref{intrep}), the structure of the 
coefficient matrix is markedly improved. Indeed, in (\ref{diffrep})
the coefficient matrix features poorly conditioned dense matrices,
save for $S_0(A)$ which is sparse and banded. Moreover, the 
derivative operators are unbounded as $N$ grows. However, in 
(\ref{intrep}) the coefficient matrix is a perturbation of the identity 
featuring banded matrices, and the integration matrices remain bounded 
as $N$ grows. 

Let $\mathcal{L}^k$ represent the matrix ---built 
from Chebyshev differentiation matri\-ces--- which represents $L^k$, 
$\mathcal{BL}^k$ the corresponding preconditioned matrix, and 
likewise for $\mathcal{P}^k$ and  $\mathcal{BP}^k$. Boundary 
conditions must be inserted into these matrices appropriately,
and we will return to this point shortly. Our goal is to compute 
and compare condition numbers for the various operators and 
boundary conditions above. Given a matrix $A$, we shall work with 
the standard condition number $\kappa(A) = \|A\|_2 \|A^{-1}\|_2$ 
belonging to the matrix 2--norm.
\begin{table}
\centering
\begin{tabular}{|l|l|l|}
\hline
Truncation & Preconditioned    &  Unpreconditioned \\
\hline
8          &  12.5839  (12.6)  &  3.7740e+03 (3.774e+03)\\
\hline
16         &  20.6559  (20.6)  &  5.5327e+04 (5.533e+04)\\
\hline
32         &  31.8782  (31.9)  &  8.4273e+05 (8.427e+05)\\
\hline
64         &  51.1595  (51.2)  &  1.3140e+07 (1.314e+07)\\
\hline
128        &  86.9027  (90.0)  &  2.0748e+08 (2.075e+08)\\
\hline
256        &  156.0339 (156.) &  3.2976e+09 (3.298e+09)\\
\hline
\end{tabular}
\\[2mm]
\caption{{\sc 2--norm condition numbers.} D--D Poisson
operator on [1,3] with $k = 3$. The numbers in parenthesis are
those listed in Tables 1 and 4
of \cite{CHHT} for the same problem. That reference's
$h_0$ condition number is the 2--norm condition number.}
\label{PoissonTable}
\end{table}

Recall that for the radiation $p,q$ boundary conditions we must 
consider consecutive blocks, since $\sin$ and $\cos$ modes of the 
same wave number are coupled. Suppose we consider the submatrix 
associated with the two consecutive blocks associated with Fourier wave 
number $k$. The unpreconditioned matrix for the helical operator 
with D--R boundary conditions is then
\begin{equation}
\left[
\begin{array}{c|c}
\begin{array}{c}
\mathcal{L}^k\\
\delta^-\\
p\delta^+
\end{array}
&
\begin{array}{c}
\mathbf{0}\\
\mathbf{0}\\
R\nu^+ + q\delta^+
\end{array}
\\
\hline
\begin{array}{c}
\mathbf{0}\\
\mathbf{0}\\
R\nu^+ + q\delta^+
\end{array}
&
\begin{array}{c}
\mathcal{L}^k\\
\delta^-\\
-p\delta^+
\end{array}
\end{array}
\right],
\end{equation}
with the understanding that the boundary conditions have been written 
over the last two rows of $\mathcal{L}^k$ in each block. However, 
as we have seen in (\ref{Ablock}), the preconditioned operator 
is
\begin{equation}
\left[
\begin{array}{c|c}
\begin{array}{c}
\delta^-\\
p\delta^+\\
\mathcal{BL}^k
\end{array}
&
\begin{array}{c}
\mathbf{0}\\
R\nu^+ + q\delta^+\\
\mathbf{0}
\end{array}
\\
\hline
\begin{array}{c}
\mathbf{0}\\
R\nu^+ + q\delta^+\\
\mathbf{0}
\end{array}
&
\begin{array}{c}
\delta^-\\
-p\delta^+\\
\mathcal{BL}^k
\end{array}
\end{array}
\right],
\label{Ablock2}
\end{equation}
now with the understanding that the boundary conditions have been
written over the first two rows of each block. 

As an example of the 
various other matrices stemming from the Poisson operator, consider 
the preconditioned matrix with Dirichlet--Neumann boundary conditions,
\begin{equation}
\left[
\begin{array}{c|c}
\begin{array}{c}
\delta^-\\
\nu^+\\
\mathcal{BP}^k
\end{array}
&
\begin{array}{c}
\mathbf{0}\\
\mathbf{0}\\
\mathbf{0}
\end{array}
\\
\hline
\begin{array}{c}
\mathbf{0}\\
\mathbf{0}\\
\mathbf{0}
\end{array}
&
\begin{array}{c}
\delta^-\\
\nu^{+}\\
\mathcal{BP}^k
\end{array}
\end{array}
\right].
\end{equation}
Here we consider this decoupled direct sum of blocks only to 
make direct comparison with the matrices above. The condition 
number of the last matrix is of course the same as the 
condition number of either of its blocks. 
\begin{table}
\centering
\begin{tabular}{|l|l|l|l|}
\hline
Truncation & Poisson (D-D) & Poisson (D-N) & Helical (D-R)\\
\hline
32         & 157.2452      & 8.0745e+03    & 9.3845e+04\\
\hline
64         & 157.9248      & 2.0750e+05    & 2.7642e+06\\
\hline
128        & 159.5918      & 6.3311e+06    & 8.4756e+07\\
\hline
256        & 164.6046      & 1.9820e+08    & 2.6537e+09\\
\hline
512        & 180.0820      & 6.2727e+09    & 8.3984e+10\\
\hline
\end{tabular}
\\[5mm]
\begin{tabular}{|l|l|l|l|}
\hline
Truncation & Poisson (D-D) & Poisson (D-N) & Helical (D-R)\\
\hline
32         & 8.8450e+05    &  6.3270e+06  &   8.6394e+05\\
\hline
64         & 1.3791e+07    &  9.8651e+07  &   1.3471e+07\\
\hline
128        & 2.1776e+08    &  1.5577e+09  &   2.1270e+08\\
\hline
256        & 3.4610e+09    &  2.4757e+10  &   3.3805e+09\\
\hline
512        & 5.5191e+10    &  3.9479e+11  &   5.3907e+10\\
\hline
\end{tabular}
\\[2mm]
\caption{{\sc 2--norm condition numbers.} Preconditioned (top) and
unpreconditioned (bottom) operators on [5,15] for
$k = 2$ and $\Omega = 0.1$.}
\label{up515}
\end{table}
\begin{table}
\centering
\begin{tabular}{|l|l|l|l|}
\hline
Truncation & Poisson (D-D) & Poisson (D-N) & Helical (D-R) \\
\hline
32       &   1.4634e+04    & 1.0922e+06    & 1.9905e+05\\
\hline
64       &   1.4640e+04    & 1.0924e+06    & 2.0635e+05\\
\hline
128      &   1.4650e+04    & 1.0949e+06    & 3.2270e+05\\
\hline
256      &   1.4673e+04    & 1.1884e+06    & 3.0785e+06\\
\hline
512      &   1.4731e+04    & 4.2550e+06    & 9.0043e+07\\
\hline
\end{tabular}
\\[5mm]
\begin{tabular}{|l|l|l|l|}
\hline
Truncation & Poisson (D-D) & Poisson (D-N) & Helical (D-R)\\
\hline
32         & 3.9154e+05    & 2.9145e+07    & 3.4677e+05\\
\hline
64         & 6.1040e+06    & 4.5437e+08    & 5.4046e+06\\
\hline
128        & 9.6378e+07    & 7.1742e+09    & 8.5335e+07\\
\hline
256        & 1.5318e+09    & 1.1402e+11    & 1.3562e+09\\
\hline
512        & 2.4426e+10    & 1.8182e+12    & 2.1627e+10\\
\hline
\end{tabular}
\\[2mm]
\caption{{\sc 2--norm condition numbers.} Preconditioned (top) and
unpreconditioned (bottom) operators on [5,150] for
$k = 2$ and $\Omega = 0.1$.}
\label{up5150}
\end{table}

For our first investigation of condition numbers we take the 
Poisson operator associated with D--D boundary conditions. We 
assume $k = 3$, a domain $[1,3]$, and consider truncations 
$N = 8,16,\cdots,256$,
where each decoupled block is $(N+1)$--by--$(N+1)$. We have
computed the conditions numbers for these truncations in
{\sc Matlab}, and recorded our results in Table \ref{PoissonTable}.
Owing to the boundary conditions which are inserted into the zeroed
rows of the preconditioned operator, we expect growth in the 2--norm
condition number for our spectral matrix representation of the
preconditioned operator as the truncation size $N$ increases. The
issue is significant for boundary conditions involving a Neumann
row vector, such as $\nu^{+}$, whose entries grow like $j^2$ with 
column location $j$. For $k = 2$ and $\Omega = 0.1$, Table \ref{up515} 
lists preconditioned (top) and unpreconditioned (bottom)
condition numbers for the short domain $[5,15]$.  Table \ref{up5150} 
lists the same numbers for the larger and more
realistic domain $[5,150]$. For the helical operator the short--domain
table shows that our preconditioning is only advantageous
initially. As the truncation size $N$ increases, the
$\tau$--conditions appear to have a dominant effect. (Since they mix
the two blocks, for the helical operator such conditions would seem
especially influential.) We loosely observe that the IPC technique 
only directly concerns the conditioning of that part of the 
coefficient matrix which stems from the operator itself, and not 
the boundary conditions {\em per se}. 
Over the whole range of truncations the
large--domain table indicates that the condition number of the 
coefficient matrix is chiefly
determined by the operator rather than the $\tau$--conditions. For
this large domain we do discern the advantage of preconditioning the
helical operator, although not as pronounced as for the Poisson
operator. As a way of obtaining a more drastic improvement in 
conditioning of the helical operator, one might explore either 
suitable equilibration tailored to the $h_r$ norms discussed in 
\cite{CHHT} or enforcement of the radiation boundary conditions via a 
penalty method\cite{H}.

\begin{figure}
\centering
\scalebox{0.75}{\includegraphics{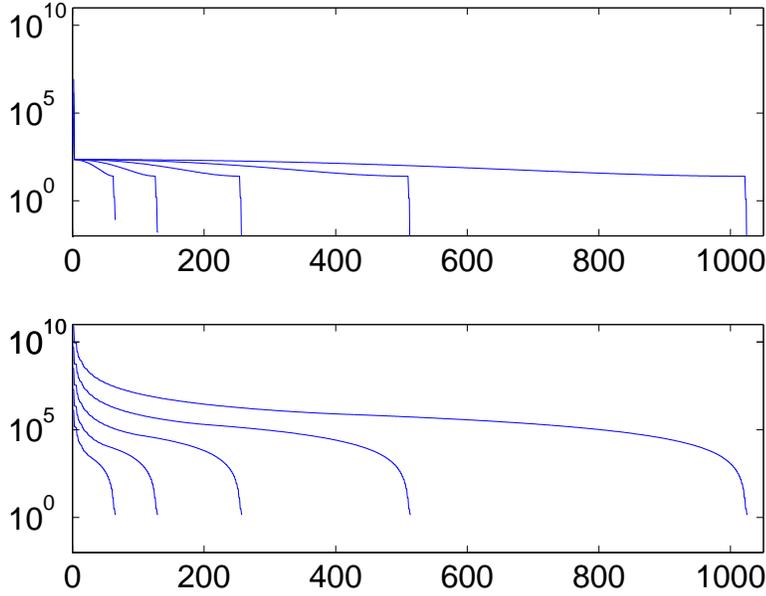}}
\caption{{\sc Singular values for last columns of Table 3.}
For operators on $[5,15]$ with $k = 2$, $\Omega = 0.1$.
}
\label{tab3singvals}
\end{figure}

\begin{figure}
\centering
\scalebox{0.75}{\includegraphics{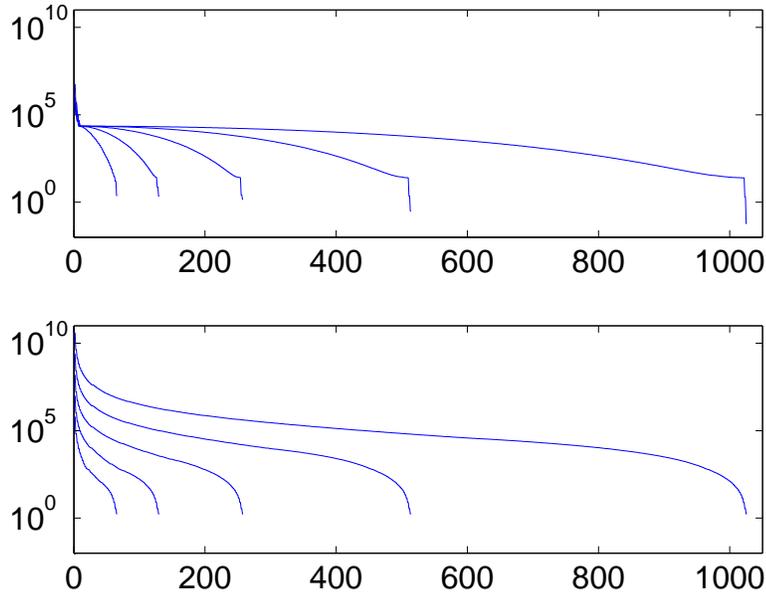}}
\caption{{\sc Singular values for last columns of Table 4.}
For operators on $[5,150]$ with $k = 2$, $\Omega = 0.1$.
}
\label{tab4singvals}
\end{figure}

In Figures \ref{tab3singvals} and \ref{tab4singvals} we plot the
singular values corresponding to the Helical (D-R) columns in
Tables \ref{up515} and \ref{up5150}, respectively. 
In Figure \ref{tab4singvals}, say,
the top plot depicts the singular value distributions corresponding
to the preconditioned matrices, for which condition numbers have
been listed in the top leftmost column of Table \ref{up5150}. The 
bottom plot shows the same data for unpreconditioned matrices, for
which condition numbers have been listed in the bottom leftmost
column of Table \ref{up5150}. In either plot, each curve corresponds to 
one entry of a column, where one should note that the number of
plotted singular values is twice the listed truncation (each matrix
consists of two modes with one block for each). Therefore, the
longest curves corresponds to the 512 truncation, and so forth.
Notice that the singular value distributions for preconditioned
matrices are more clustered.

\subsection{Numerical results for the linear and nonlinear 2d HRWE}
\label{subsec:2dsims} Throughout this section we examine the 2d model, 
and, unless explicitly stated otherwise, all matrix inversions have 
been carried out with {\tt dgesv}. Those inversions carried out with
{\tt dgesvx} will be clearly indicated.

Consider the linear equation
\begin{equation}
L\psi = Q \frac{\delta(r-x_H)}{x_H}
[\delta (\varphi-\pi) - \delta (\varphi)].
\label{deltasource}
\end{equation}
where $L$ is the HRWE operator and the inhomogeneity is comprised 
of two equal but opposite strength $\delta$--functions, one located 
at $(x,y) = (x_H,0)$ and the other at $(x,y) = (-x_H,0)$. Assuming 
that exact outgoing radiation conditions are placed at 
$r = R = r_\mathrm{max}$, we may express the exact solution 
to (\ref{deltasource}) as the following infinite 
series\cite{WheKrivPri}:
\begin{equation}
\psi(r,\varphi) =
-\sum_{m}
Q J_m(m\Omega r_<)
\big[N_m(m\Omega r_>)
\cos(m\varphi)+J_m(m\Omega r_>)\sin(m\varphi)\big]
\label{outgoingseries}
\end{equation}
where the sum is over $m = 1,3,5,\cdots$ and $r_>$ is the larger 
of $x_H$ and $r$ and similarly for $r_<$. In what follows we always 
take $x_H = 2$, $\Omega = 0.1$, and $Q = 1$. Although we shall not 
carefully study the pointwise convergence of
the series (\ref{outgoingseries}), we note that it is poorly convergent
whenever $r$ is close to $x_H$. In using (\ref{outgoingseries}) to
generate approximations to field values $\psi$ for the experiments
below, we have at times used 30 digit precision in {\sc Maple} and
included thousands of terms in order to sum the series with double
precision accuracy.

In order to make a direct comparison between the series 
(\ref{outgoingseries}) and numerical solutions, we now consider 
the following three {\em linear} and {\em homogeneous} problems.
\begin{figure}
\centering
\scalebox{0.39}{\includegraphics{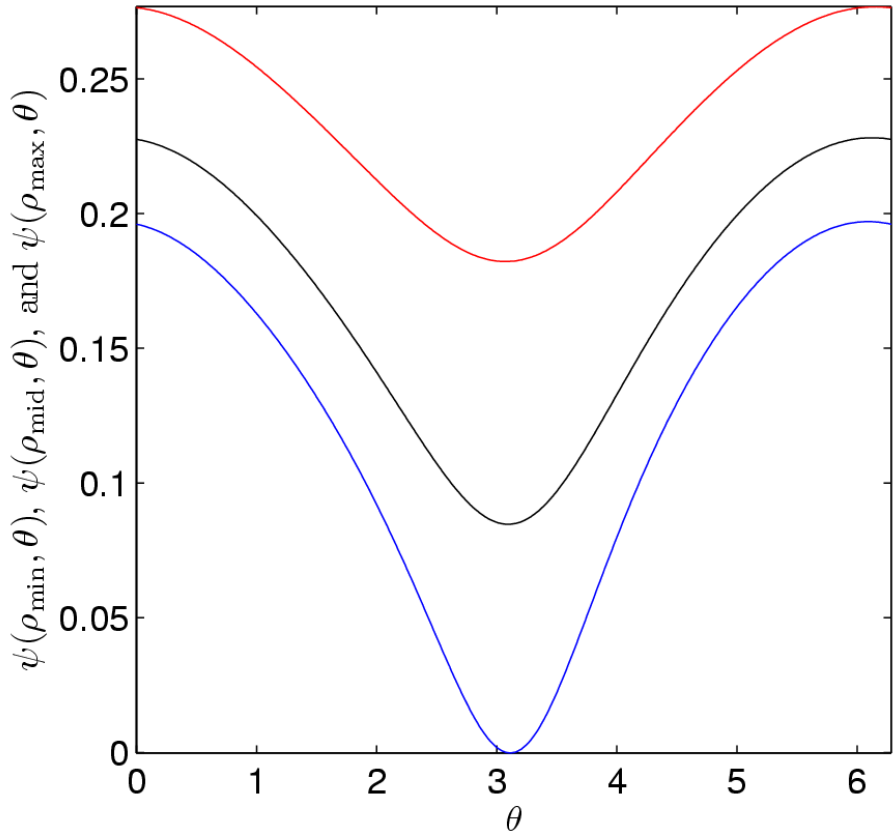}}
\caption{{\sc Constant--$\rho$ sections of
$\psi(\rho,\theta)$.} The figure depicts the
solution (\ref{outgoingseries}) expressed in terms of the
local polar coordinates $(\rho,\theta)$ about the hole. The
top curve corresponds to $\rho_\mathrm{min} = 1$, the middle to
 $\rho_\mathrm{mid} = 1.5$, and the bottom to
$\rho_\mathrm{max} = 2$.}
\label{ring}
\end{figure}

\begin{itemize}
\item[i.]{{\em Elliptic problem on inner annulus.} Let $\rho$ 
represent the radial coordinate as measured from the source 
point $(x_H,0)$, so that the inner annulus corresponds to 
$\rho_\mathrm{min} \leq \rho \leq \rho_\mathrm{max}$. We
consider the homogeneous problem $L\psi = 0$, using the
series (\ref{outgoingseries}) to place Dirichlet boundary 
conditions at $\rho = \rho_\mathrm{min}$ and 
$\rho = \rho_\mathrm{max}$.}

\item[ii.]{{\em Mixed problem on outer annulus.}}
We consider $L\psi = 0$ on the outer annulus corresponding
to $r_\mathrm{min} = \varepsilon \leq r \leq R = r_\mathrm{max}$,
using the series (\ref{outgoingseries}) to place a Dirichlet 
boundary at $r = \varepsilon$ and adopting exact outgoing
conditions at $r = R$. In this setting, we assume $\varepsilon > 
x_H$ (where $\varepsilon$ need not be small). Again, the light
circle lies between $\varepsilon$ and $R$.

\item[iii.]{{\em Mixed problem on two center domain.} In this setting
we consider the linear problem $L\psi = 0$ on the two center domain 
depicted in Fig.~\ref{domaindecomp}.
We use the series 
(\ref{outgoingseries}) to place Dirichlet boundary conditions
at $\rho = \rho_\mathrm{min}$ (for both inner ``holes'') and
adopt exact outgoing conditions at $r = R$.}
\end{itemize}
In each case, we will generate a numerical solution, and
then compare it with the series via various choices of error measure. 
Our numerical experiments for cases (i) and (ii) are meant to 
facilitate better parameter choices for scenario (iii) experiments.

\subsubsection{Linear elliptic problem on inner annulus.}
For the numerical experiment in (i) above, we use the series 
to generate Dirichlet data at $\rho_\mathrm{min} = 1$ and $\rho_\mathrm{max} 
= 2$. With a double Chebyshev--Fourier expansion and the described 
integration preconditioning, we  numerically solve $L\psi = 0$ on this 
inner annulus. From the spectral solution we then generate numerical 
values for 
$\psi(\rho_\mathrm{mid},\theta)$, where $\rho_\mathrm{mid} = 1.5$ and 
$\theta$ is sampled on a fine grid. This $\rho_\mathrm{mid}$ profile is then 
compared with the analogous profile given by the exact series 
(\ref{outgoingseries}). All profiles are depicted in Fig.~\ref{ring}, 
and Table \ref{innerannuluserrors} lists the associated errors for the
$\rho_\mathrm{mid}$ profile, where $N_H$ denotes the number of radial
Chebyshev elements, and $M_H$ the number of angular Fourier elements.
\begin{table}
\centering
\begin{tabular}{|c||c|c|c|c|}
\hline
               &   absolute      &  relative       &  absolute       &  relative     \\
$N_H$, $M_H$   &   sup error     &  sup error      &  rms error      &  rms error    \\
\hline
8,   16        &   1.091e$-$05   &   4.782e$-$05   &   5.159e$-$06   &   2.964e$-$05 \\
\hline
12,  32        &   2.551e$-$09   &   1.118e$-$08   &   1.226e$-$09   &   7.040e$-$09 \\
\hline
16,  48        &   7.646e$-$13   &   3.352e$-$12   &   3.697e$-$13   &   2.124e$-$12 \\
\hline
20,  64        &   3.192e$-$16   &   1.399e$-$15   &   1.352e$-$16   &   7.766e$-$16 \\
\hline
\end{tabular}
\\[2mm]
\caption{{\sc Errors for the inner annulus problem.}}
\label{innerannuluserrors}
\end{table}

In this table, as in all remaining tables of the paper, the meaning of each 
of the error measures is as follows. The absolute sup error is simply the 
magnitude of the largest error found in the computed solution. For our 
spectral solutions this means that the solutions must be computed and compared on 
an evaluation grid. For Table \ref{innerannuluserrors} the grid was taken to 
be 1024 evenly spaced points of $\theta$, at $\rho_{\rm mid}$. The relative 
sup error is the absolute sup error divided by the maximum value found on 
the evaluation grid. The absolute rms error is found by taking the sum of
the squares of all absolute errors on the evaluation grid, then dividing 
that sum by the number of terms that have been added, and finally by taking the 
square root. The relative rms error is the absolute rms error divided by 
the rms value of the solution on the evaluation grid. To evaluate the 
exact series on the evaluation grid, we have first evaluated it to high 
accuracy with {\sc Maple} on a (uniformly spaced) Fourier grid of 128 points, 
subsequently using the corresponding Fourier interpolation onto the finer 
grid. Direct evaluation of the series on the 1024 point grid proved
impractical.
\begin{table}
\centering
\begin{tabular}{|c||c|c|c|c|}
\hline
               &   absolute      &   relative      &   absolute      &   relative     \\
$N_A$, $M_A$   &   sup error     &   sup error     &   rms error     &   rms error    \\
\hline
20,  11        &   5.433e$-$04   &   3.114e$-$03   &   3.223e$-$05   &   9.171e$-$01 \\
\hline
38,  31        &   1.045e$-$07   &   5.991e$-$07   &   3.001e$-$09   &   2.142e$-$04 \\
\hline
56,  51        &   3.028e$-$11   &   1.736e$-$10   &   6.767e$-$13   &   4.831e$-$08 \\
\hline
80,  81        &   6.661e$-$16   &   3.819e$-$15   &   8.937e$-$17   &   6.380e$-$12 \\
\hline
\end{tabular}
\\[2mm]
\begin{tabular}{|c||c|c|c|c|}
\hline
               &   absolute      &   relative      &   absolute      &   relative    \\
$N_A$, $M_A$   &   sup error     &   sup error     &   rms error     &   rms error   \\
\hline
32,  11        &   5.433e$-$04   &   3.114e$-$03   &   3.885e$-$05   &   9.979e$-$01 \\
\hline
76,  31        &   1.045e$-$07   &   5.991e$-$07   &   5.601e$-$09   &   2.223e$-$03 \\
\hline
124,  51       &   3.028e$-$11   &   1.736e$-$10   &   8.612e$-$13   &   3.419e$-$07 \\
\hline
200,  81       &   9.853e$-$16   &   5.649e$-$15   &   2.981e$-$16   &   1.183e$-$10 \\
\hline
\end{tabular}
\\[2mm]
\caption{{\sc Errors for the outer annulus problem.} The top table corresponds to
$R = 50$ and the bottom to $R = 150$. These numbers have been extracted from a
1024 $\times$ 1024 uniform grid.}
\label{annulustable}
\end{table}
\begin{figure}
\centering
\scalebox{0.48}{\includegraphics{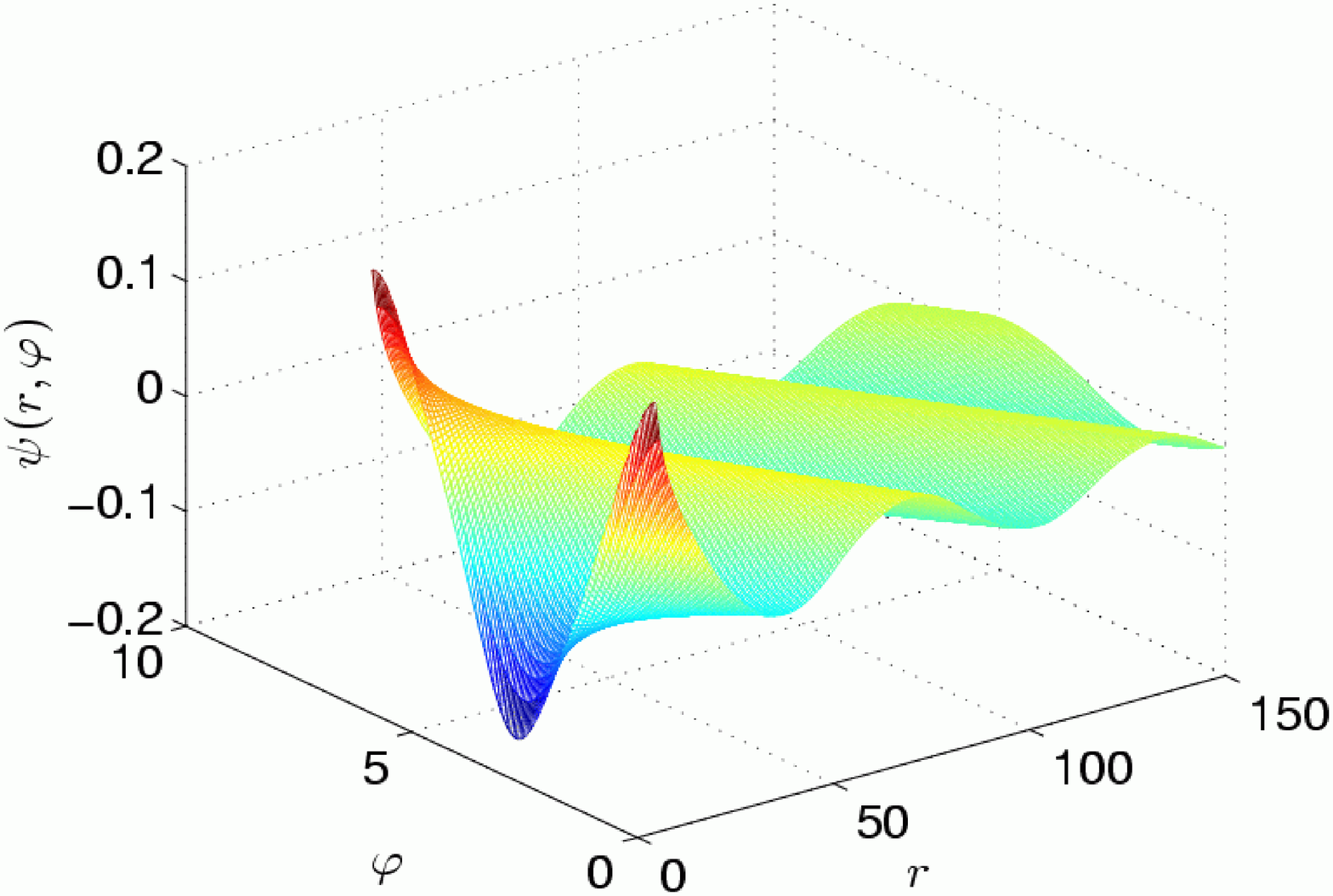}}
\scalebox{0.478}{\includegraphics{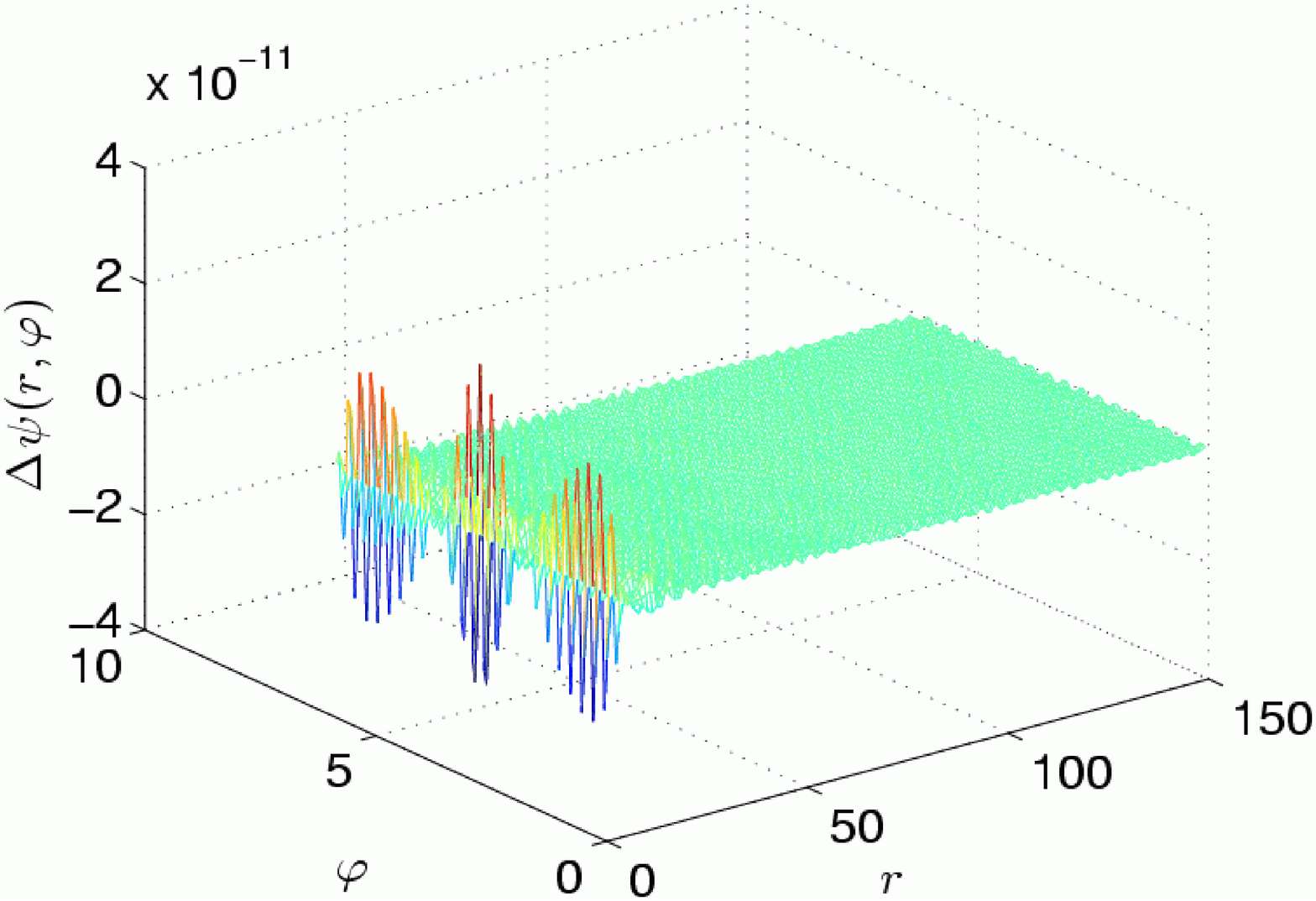}}
\caption{{\sc Outer annulus problem.} The plots depict the numerical
solution and error for $N_A = 124$ and $M_A = 51$.}
\label{annuluserr}
\end{figure}

\subsubsection{Linear mixed problem on outer annulus}\label{outerannulusprob}
For this problem, 
in the region $A$ of Fig.~\ref{domaindecomp},
we choose $\varepsilon = 4.5$ and either $R = 50$ or $R = 150$. 
Since $\Omega = 0.1$, the light circle always lies within the 
corresponding annulus. Using (\ref{outgoingseries}) to generate 
Dirichlet data on the circle $r = \varepsilon$ is not problematic, 
since $4.5 > 2$ and so the series converges reasonably well. We 
have used the {\tt FORTRAN} routines {\tt RJBESL} and {\tt RYBESL} 
from {\tt netlib} to build up the series (and also {\sc Matlab}'s
Bessel routines for error analysis). In Table \ref{annulustable}
we list error measures in the outer annulus for various choices
of $N_A$ (number of radial Chebyshev elements)
and $M_A$ (number of Fourier elements). In each case, for a fixed
choice of $M_A$ we have experimented to (roughly) find the
smallest value $N_A$ which achieves the best possible errors.
Figure \ref{annuluserr} depicts the numerical solution
$\psi(r,\varphi)$ and the error $\Delta\psi(r,\varphi)$, the 
difference between the numerical and series solutions, 
corresponding to the bottom table's second
to last line. Notice that most of the error is concentrated near 
the inner boundary $r = \varepsilon$, and this region determines
the supremum errors in the tables. For a fixed particular number 
of angular Fourier elements, these values are, save for the last 
line, the same for both tables.

We remark that owing to the quadratic growth inherent in the our radiation 
boundary conditions (based on Neumann vectors $\nu^+$), 80 or 200 radial 
elements is likely excessive. We should prefer to further divide $A$ into 
sub--annuli all glued together, with fewer Chebyshev elements taken on 
each sub--annulus. In future 3d work with numerical shells we plan to 
explore this possibility.

\subsubsection{Linear mixed problem on two center domain: 
iteration between the elliptic region and outer annulus}
This numerical experiment tests the full multidomain spectral code. 
We use the series (\ref{outgoingseries}) summed in extended precision
with {\sc Maple} to place Dirichlet conditions on the inner holes (in
practice, owing to the symmetry of the problem, on one inner hole
$H$). We place radiation conditions at the outer boundary $r = R$ of
the outer annulus. We take the hole $H$ and all rectangles 1--8 of
Fig.~\ref{domaindecomp} as one glued domain, confined within a large
square with side $2L = 10$. The outer annulus $A$ is
treated as a separate domain, and we iterate between solves on the
hole--plus--rectangles domain and the outer annulus.  More precisely,
we first solve the hole--plus--rectangles configuration using the
series (\ref{outgoingseries}) to place Dirichlet conditions on the
inner boundary $\rho = \rho_\mathrm{min}$ of the hole and adopting
Dirichlet zero conditions on the free edges of rectangles which
overlap the outer annulus $A$. This multidomain numerical solution
then provides inner Dirichlet conditions at $r = \varepsilon = 4.5$ 
for the solve on the outer annulus. The solution for the outer annulus, 
in turn, provides a numerical solution from which we obtain corrected
Dirichlet conditions for the relevant free edges of rectangles. While
this iteration between solves on the hole--plus--rectangles domain and
outer annulus is under way, we monitor the rms error between the inner
profile $\psi(\varepsilon,\varphi)$ from the outer annulus at the 
current and previous iterate. The iteration is stopped once this 
inner boundary rms error ceases to decrease, typically after 100 or so 
iterations. We could of course glue the 
outer annulus to the hole--plus--rectangles domain, and indeed we do so 
later when examining the nonlinear model problem, but here we use the 
described iteration, since we wish to demonstrate its stability. Table
\ref{twocentertable} lists error measures associated with this
experiment with the outer boundary taken as $R = 50$.  In the table 
the choices for $N_H$, $M_H$ and $N_A$, $M_A$ stem from our earlier 
experiments on annulus domains. For simplicity, we have chosen the 
same numbers $N$, $M$ for the elements associated with the double 
Chebyshev expansion on each of the rectangular domains. For the 
table's last line, we have used {\tt dgesvx} in lieu of {\tt dgesv},
for the hole--plus--rectangles domain only. Without
the extra accuracy afforded by {\tt dgesvx} for the elliptic region, 
the numbers in the last line would be only marginally better than 
those in the next to last. See the final paragraph in  
Sec.~\ref{subsubsec:glueRecs} for further germane comments.
\begin{table}
\centering
\begin{tabular}{|c|c|c||c|c|c|c|}
\hline
             &              &            &  absolute     &  relative     &  absolute     &  relative   \\
$N_H$, $M_H$ & $N$, $M$ & $N_A$, $M_A$   &  sup error    &  sup error    &  rms error    &  rms error  \\
\hline
 8, 16       & 8,8          & 20, 11     &  4.540e$-$04  & 2.600e$-$03   & 5.362e$-$05   & 9.674e$-$01\\
\hline
 12, 32      & 20, 20       & 38, 31     &  9.908e$-$08  & 5.680e$-$07   & 2.988e$-$09   & 2.133e$-$04\\
\hline
 16, 48      & 30, 30       & 56, 51     &  3.023e$-$11  & 1.733e$-$10   & 7.907e$-$13   & 5.645e$-$08\\
\hline
 20, 64      & 38, 38       & 80, 81     &  1.402e$-$15  & 8.035e$-$15   & 2.291e$-$16   & 1.636e$-$11\\
\hline
\end{tabular}
\\[2mm]
\caption{{\sc Errors for the two center domain problem.} These error measures
have been extracted from the outer annulus only (again with a $1024\times 1024$
uniformly spaced grid), although the numerical solution is generated on the
entire two center domain. The last line of this table has been obtained with
{\tt dgesvx} (see text). Compare this table with the top table in Table
\ref{annulustable}.}
\label{twocentertable}
\end{table}
\begin{figure}
\centering
\scalebox{0.455}{\includegraphics{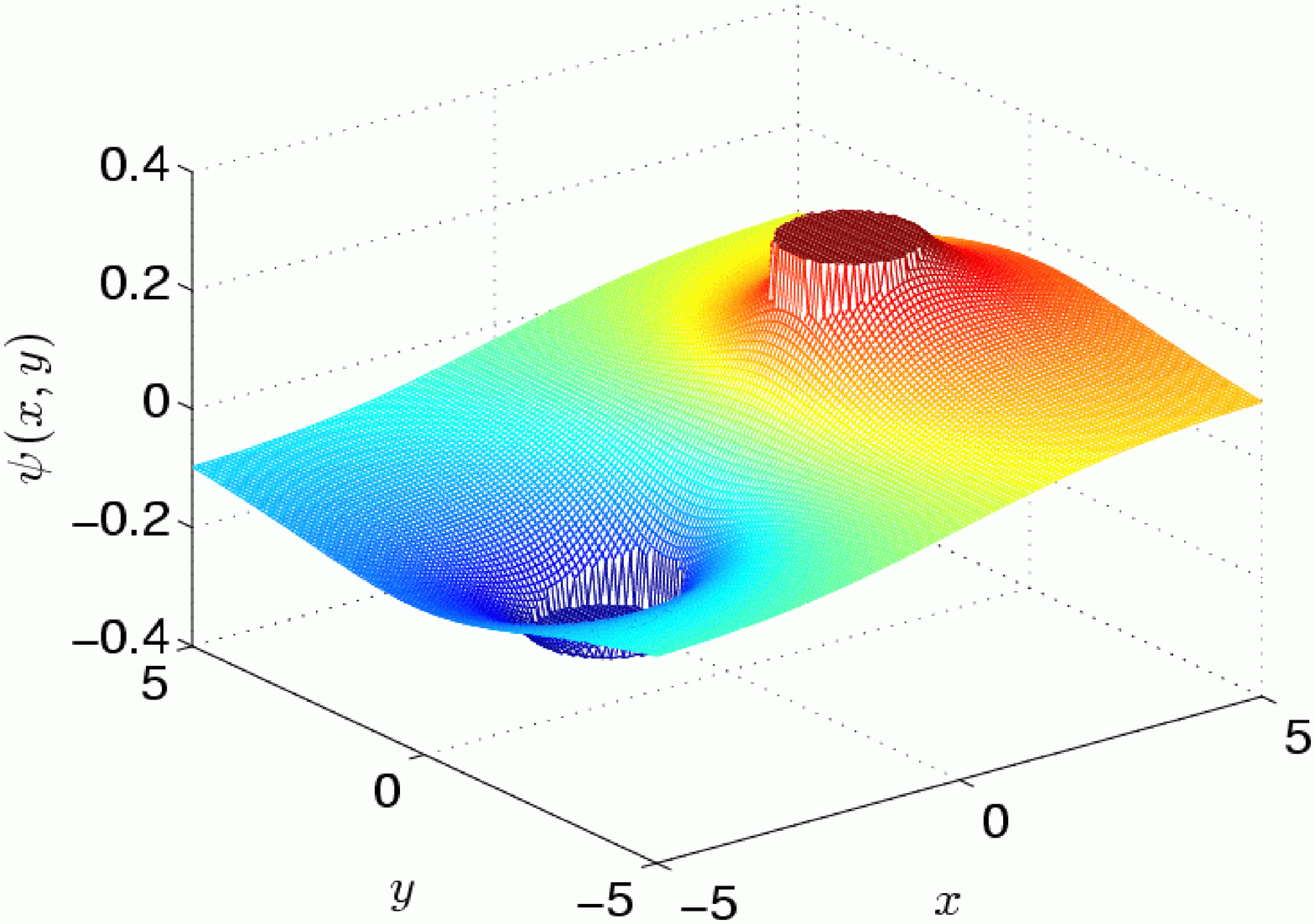}}
\scalebox{0.4475}{\includegraphics{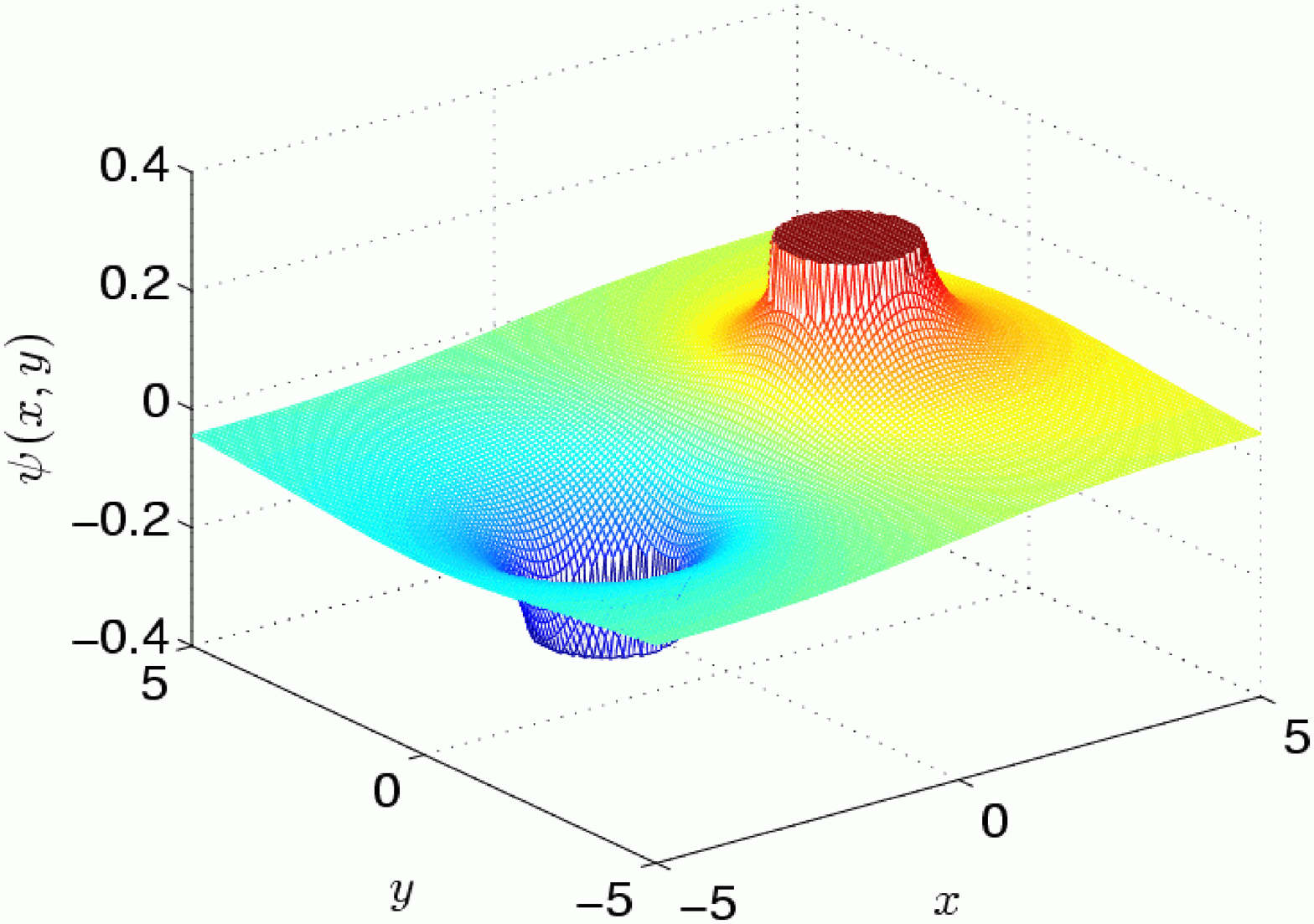}}
\caption{{\sc Comparison of solutions}. The plots
depicts numerical solutions for the linear (top)
and nonlinear (bottom) models. In each case the
solution is not known in the immediate neighborhood
of the holes, whence on these neighborhoods we have
set the solution to a constant value in order to
achieve nice plots.}
\label{compareltonl}
\end{figure}

\subsubsection{Nonlinear mixed problem on two center domain}\label{nonlinmod2c}

For our final experiment in 2d, we examine the homogeneous ($g = 0$) nonlinear 
equation (\ref{nonlin2dHRWE}). For nonlinear solves we have always chosen to 
work with the fully glued two center domain ($H$ + all rectangles 1--8 + $A$).
Our numerical experiment with the nonlinear model is to repeat the mixed boundary 
value problem carried out for the linear model on the full two center domain. In 
particular, we use the same Dirichlet conditions, from (\ref{outgoingseries}), 
at $\rho=\rho_{\rm min}$. A major difference is that now we solve simultaneously 
for the unknowns in all subdomains rather than by back and forth iteration with 
the outer annulus as we did for the linear problem. Since there is no exact 
solution available for comparison, we simply compare solutions with different 
resolutions. As before, the comparison is carried out using output associated with 
the outer annulus on a 1024 $\times$ 1024 uniform grid. We take $R = 50$, 
$\eta = -10$, and $\psi_0 = 0.25$. With a numerical solution to the nonlinear 
problem, we may naively compute its ``dipole strength'' via Fourier transform of 
the solution restricted to the outer boundary $r = R$. The nonlinear term 
corresponding to our parameter choices results in a dipole strength which differs 
from dipole strength for the linear model ($\eta = 0$) by about 18 percent, and 
this difference measures the nonlinear term's strength. Figure \ref{compareltonl} 
depicts solutions for both the linear and nonlinear models 
in the vicinity of the holes, showing that even to the eye these solutions
differ. Error measures are collected in Table \ref{nonlintwocentertable}. 

\subsection{Preliminary results for the linear 3d HRWE}\label{3dsim}

We briefly describe two numerical experiments involving the 3d HRWE and meant
to suggest that a multidomain spectral approach is viable for 3d PSW approximations, 
at least insofar as the post--Minkowski scenario is concerned and possibly beyond. 
The first experiment involves the 3d mixed--problem on an outer spherical shell, 
analogous to the outer annulus problem studied in Sec.~\ref{outerannulusprob}. The 
second experiment is a crude two--domain model consisting of an inner cube (lying 
within the elliptic region and on which we place explicit sources) and an 
outer spherical shell (intersecting the type--changing cylinder). This single 
cube serves as a crude substitute for all the subdomains depicted in 
Fig.~\ref{domaindecomp3} (as well as the inner spherical shell not depicted). We
would prefer an experiment in which a single cylinder replaced all these 
subdomains, but as yet have not developed a library of spectral routines for 
solid cylinders. Our simple two--domain model indicates that iteration between 
an inner elliptic region and an outer spherical shell is stable. 
Such an iteration also appears to be quite stable in the 2d setting.
\begin{table}
\centering
\begin{tabular}{|c|c|c||c|c|c|c|}
\hline

             &              &           &   absolute        &   relative        &   absolute     &   relative    \\
$N_H$, $M_H$ & $N$, $M$  & $N_A$, $M_A$ &   sup error       &   sup error       &   rms error    &   rms error   \\
\hline
 8, 16       & 8, 8      & 20, 11       &   1.096e$-$03     &   1.211e$-$02     &   1.371e$-$04  &   9.976e$-$01 \\
\hline
 14, 36      & 20, 20    & 42, 39       &   1.152e$-$07     &   1.273e$-$06     &   1.140e$-$08  &   1.209e$-$03 \\
\hline
 18, 54      & 30, 30    & 64, 65       &   3.130e$-$10     &   3.459e$-$09     &   5.991e$-$12  &   6.354e$-$07 \\
\hline
\end{tabular}
\\[2mm]
\caption{{\sc Errors for the nonlinear two center domain problem.} As with
the linear model experiment, these error measures are extracted from the
outer annulus only. Error measures for the first two lines are taken
with respect to the numerical solution corresponding to the next line.
Error measures in the last line are taken with respect to the numerical
solution which corresponds to incrementing all parameters by 2.}
\label{nonlintwocentertable}
\end{table}

\subsubsection{Mixed problem on outer shell}
\label{subsubsec:mixedshell}
For the first experiment we define the outer spherical shell via 
$4 = \varepsilon \leq r \leq R = 50$, and consider the following 
exact series solution to the homogeneous 3d HRWE:
\begin{align}
\psi(r,\theta,\varphi) 
= & \sum_{\ell=0}^{\ell_\mathrm{max}} c_{\ell 0}
Y_{\ell 0}(\theta,0)\frac{a^\ell}{r^{\ell+1}}
+ \sum_{\ell=1}^{\ell_\mathrm{max}} \sum_{m=1}^\ell 
   c_{\ell m} Y_{\ell m}(\theta,0) j_\ell(m\Omega a)
\nonumber \\
& \times
         \big[j_\ell(m\Omega r)\sin(m\varphi)+
              n_\ell(m\Omega r)\cos(m\varphi)\big],
\label{sphericalbesselseries1}
\end{align}
here with $r > a = 1$ and $\Omega = 0.2$. The $c_{\ell m}$ are
random expansion coefficients, obeying $|c_{\ell m}| \leq 1$ and drawn from 
a uniform distribution, $Y_{\ell m}(\theta,\varphi)$ is a scalar 
spherical harmonic, and $j_\ell(z)$ and $n_\ell(z)$ are spherical Bessel 
functions. Notice that $\varepsilon < |\Omega|^{-1}$, whence the inner 
surface of the shell lies within the elliptic region, as discussed 
in Sec.~\ref{subsec:prelim3dHRWE}. The form of this series is motivated 
by the rather more physical series\cite{BBP2006}
\begin{align}
\psi(r,\theta,\varphi)
= & -2K\sum_{\ell=0}^{\infty}
        \frac{1}{2\ell+1}
                   Y_{\ell 0}(\pi/2,0)
                   Y_{\ell 0}(\theta,0)
                   \frac{a^\ell}{r^{\ell+1}}
\nonumber \\
& + 4 K\Omega \sum_{\ell=2}^{\infty}
      \sum_{m=2,4,6,\cdots}
      m Y_{\ell m}(\pi/2,0)
        Y_{\ell m}(\theta,0)
\nonumber \\
& \times j_\ell(m\Omega a)
         \big[j_\ell(m\Omega r)\sin(m\varphi)+
              n_\ell(m\Omega r)\cos(m\varphi)\big]
\label{sphericalbesselseries2}
\end{align}
corresponding to two equal point charges, that is
\begin{equation}
g(r,\theta,\varphi) = 
K \frac{\delta(r-a)}{a^2}\delta(\cos\theta)\big[\delta(\varphi)
+\delta(\varphi+\pi)\big]
\end{equation}
in Eq.~(\ref{helical30}). The choice here of like signs for the point 
sources would make the problem similar to the gravitational problem, with 
radiation dominated by the quadrupole mode. (In the 2d case we chose equal 
and opposite signs to avoid the bothersome logarithmic 2d monopole.) 
We could of course base our test on a truncation of 
(\ref{sphericalbesselseries2}), but will instead work with 
(\ref{sphericalbesselseries1}) in order to test all modes defined for a 
given choice of $\ell_\mathrm{max}$.

We use the series (\ref{sphericalbesselseries1}) to seed inner boundary 
conditions at $r = \varepsilon$, which would
also be possible for (\ref{sphericalbesselseries2}), since the locations 
$r = a$, $\theta = \pi/2$, $\varphi = 0,\pi$ do not lie in the shell, 
and so the series is convergent for $r > a$. This requires a 
spherical--harmonic decomposition, which we perform with the NCAR routines 
SPHEREPACK by Adams and Swarztrauber\cite{NCAR}. For this example, we have 
coded our spectral representation of the 3d HRWE as a matrix--vector multiply. 
As we do not explicitly form the matrix, we solve the resulting linear system 
using the iterative method GMRES\cite{CERFACS}. Our matrix--vector multiply 
involves integration preconditioning in the radial variable only. We use the series 
above with $\ell_\mathrm{max} = 7$, which corresponds to $1+3+5+7+9+11+13+15 = 
64$ modes. The capacity to perform spectral analysis and synthesis (transform and 
inverse transform) then requires that a corresponding physical grid has an angular
resolution of at least $N_\theta \times N_\varphi = 8 \times 15 = 120$ 
points. In anticipation of using such analysis and synthesis when dealing 
with nonlinearities, we therefore work with a set of unknowns corresponding 
to $N_r\times N_\theta\times N_\varphi = 60 \times 8 \times 15 = 7200$, 
although $60 \times (120-64) = 3360$ should be zero. Since only the remaining
$60\times 64 = 3840$ unknowns should be nonzero, as our linear solver we use 
GMRES, taking the number of iterations $M_\mathrm{itr} \leq 3840$. As an example, 
with $M_\mathrm{itr} = 2000$ we produce a numerical solution from which we plot 
and analyze various cross sections. Over the equatorial plane the numerical 
solution has a supremum error of $1.194 \times 10^{-9}$. For $M_\mathrm{itr} = 
1600$ the corresponding supremum error is $1.038\times 10^{-6}$.
\begin{figure}
\centering
\scalebox{0.46}{\includegraphics{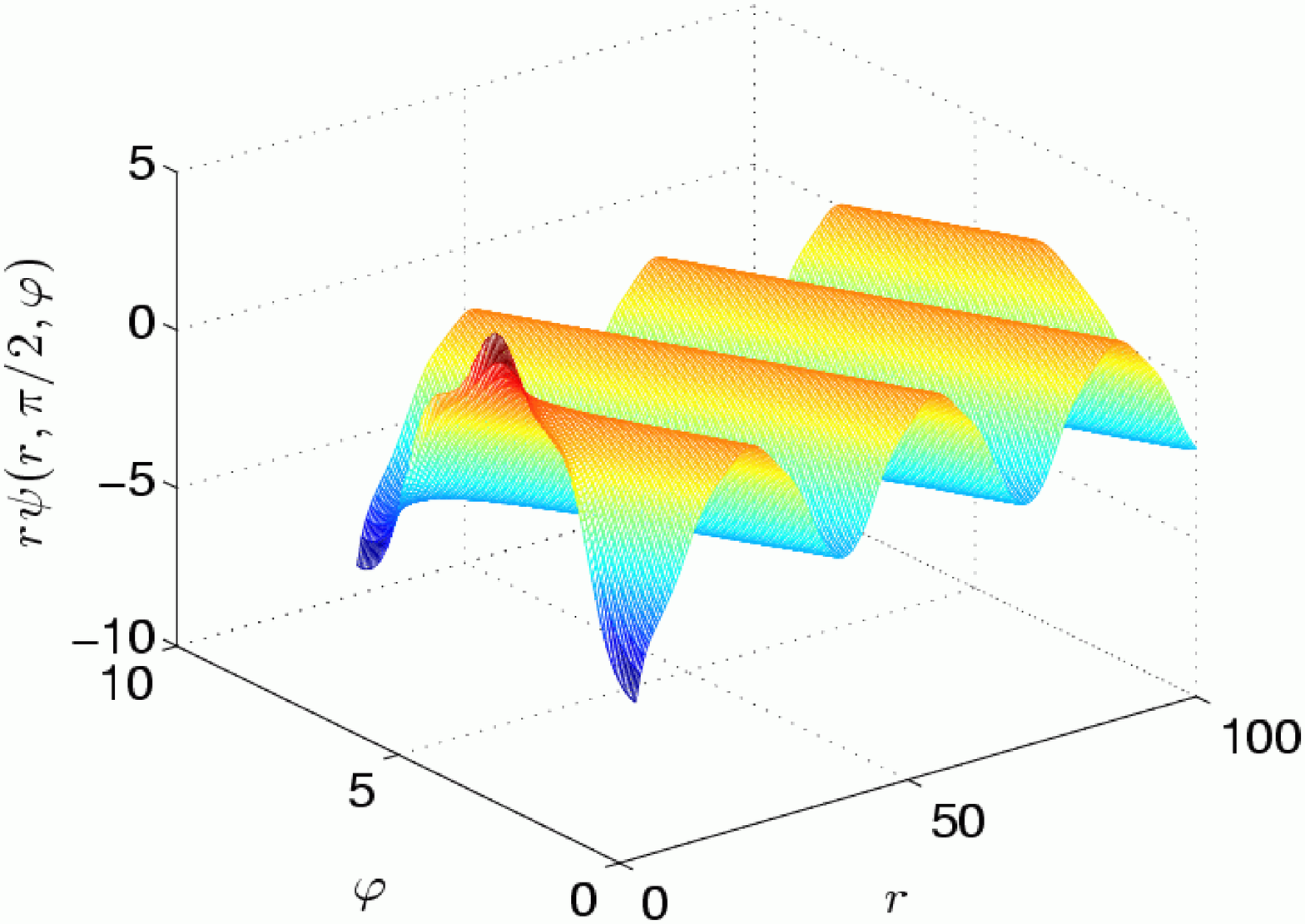}}
\scalebox{0.4525}{\includegraphics{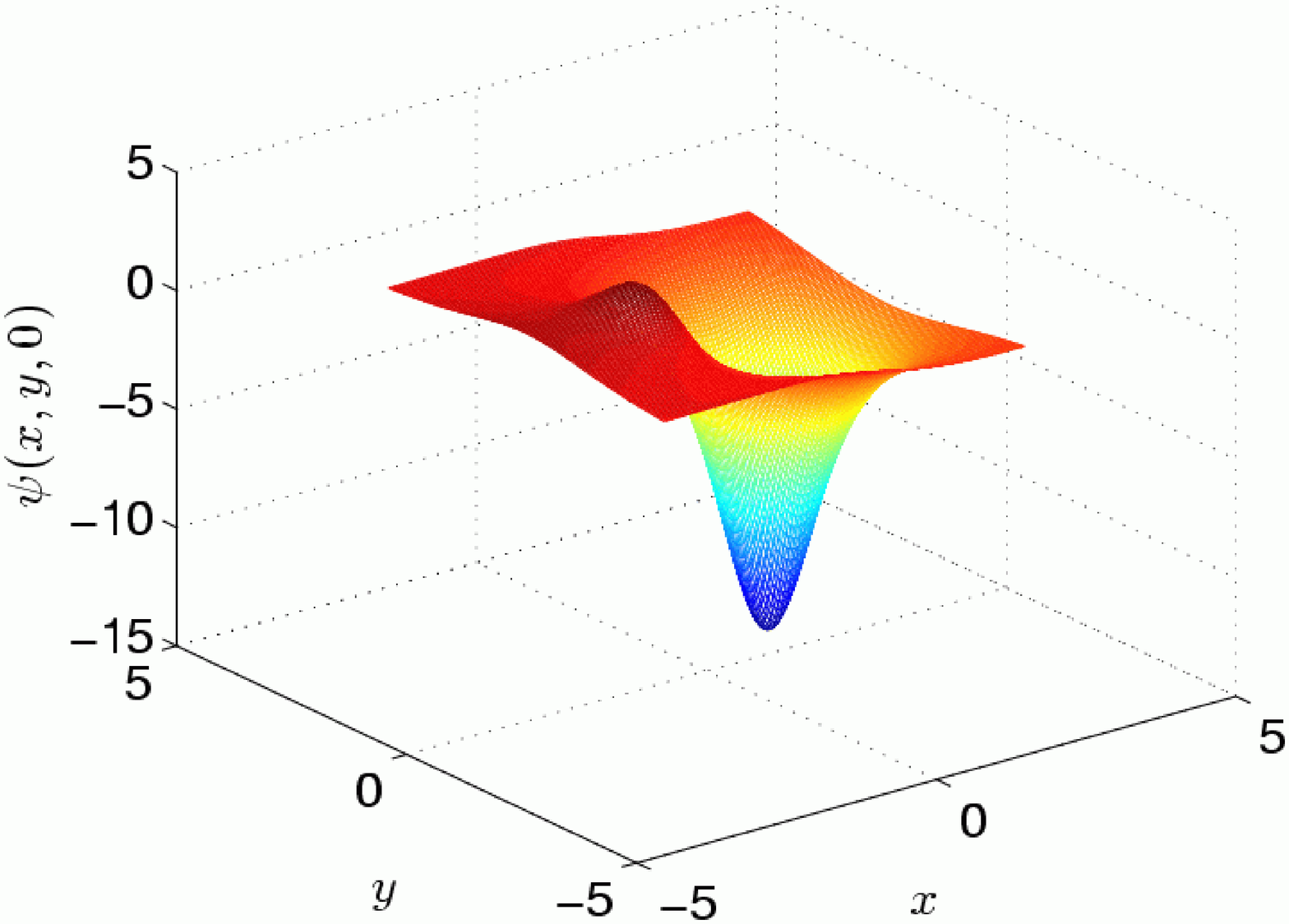}}
\caption{{\sc Cross sections of the 3d numerical solution.}
The top plot depicts the equatorial cross section on the outer
shell, with multiplication by $r$ enhancing $1/r$ fall off.
The bottom plot depicts the equatorial cross section on the
inner cube.}
\label{cubeshellslice}
\end{figure}

\subsubsection{Mixed problem on two coupled domains}
As the second experiment, we consider essentially the same problem described 
in Ref.~\cite{Andradeetal}, that is the linear inhomogeneous 3d HRWE of
(\ref{helical30}) with two Gaussian sources,
\begin{equation}
g(r,\theta,\varphi) =
Q e^{-\kappa\big[(x-x_H)^2+(y-y_H)^2+z^2\big]}
+\mu Qe^{-\kappa\big[(x+x_H)^2+(y+y_H)^2+z^2\big]}.
\end{equation}
Our parameter choices are $Q = 100$, $\kappa = 3$, with the location of the 
``hole'' (source) at $(x_H,y_H) = (1,0)$ in the $z=0$ plane. Furthermore, we 
choose $\mu = -0.25$ in order to ensure that the stability of our scheme is 
not predicated on symmetry. We determine the HRWE operator by $\Omega = 0.2$. 
Our two basic subdomains are an inner cube and an outer shell, 
respectively determined by 
$[x_\mathrm{min},x_\mathrm{max}] =
[y_\mathrm{min},y_\mathrm{max}] = [z_\mathrm{min},z_\mathrm{max}] =
[-3,3]$ and $[\varepsilon,R] = [2.8,100]$. 

To solve this problem numerically, we work with the two domains as decoupled. 
On the cube we use a pseudospectral method (point values as unknowns) based 
on GMRES, with the derivatives in the HRWE operator coded as a matrix--vector 
multiply. We currently employ no preconditioning whatsoever on the cube in 
our simple model, and so are limited in what resolutions we can obtain. For
the experiment we have taken a $N_x\times N_y \times N_z = 30\times 30\times 
30$ truncation. For the outer shell, we again use GMRES but with the purely 
spectral method just described in Sec.~\ref{subsubsec:mixedshell}. 
Therefore, we have adopted 
a hybrid pseudospectral/spectral approach. Our resolution in the annulus 
corresponds to a grid size of $N_r \times N_\theta \times N_\varphi = 62 
\times 5 \times 9$. Our GMRES--based experiment requires less than 5 percent 
of the combined storage required to perform an individual direct solve for 
each domain, although it does not achieve high accuracy.

Starting with the cube, we iterate between the solves. That is to say, we 
first solve the problem on the cube with Dirichlet--zero boundary conditions. 
Via the global interpolation then available from the cube solution, we seed 
inner Dirichlet data $\psi|_\mathrm{\varepsilon}$ for the shell solve. Since
we are solving the 3d HRWE on a closed ball (cube plus shell) with no inner 
boundaries, the solution is only determined up to a free constant. We fix
this constant by setting to zero the $\ell = 0$ mode of 
$\psi|_\mathrm{\varepsilon}$. The solution on the shell then provides 
corrected Dirichlet data for another cube solve, and so on. The iteration
proceeds until the rms error between successive seed data sets 
$\psi|_\mathrm{\varepsilon}$ ceases to decrease (about 60 iterations).
Figure \ref{cubeshellslice} shows the resulting numerical solution. For 
this low resolution experiment, the depicted outer shell solution has an 
absolute supremum error (as measured against a second numerical solution 
obtained with slightly more resolution on both the cube and shell, but 
with the same number of angular modes) measuring $1.478\times 10^{-2}$ 
over the equatorial plane.
%
%
Although we have not considered
large values of $\Omega$, we note that the iteration appears to be 
completely stable and convergent. An instability could set in at higher 
resolutions. Nevertheless, we find the results from this 
model 3d code as well as similar iteration in the 2d setting encouraging. 

\section{Summary, conclusions, and outlook}

The motivation for this work arose in problems of gravitational waves 
and black holes.  In a late stage of inspiral the interaction of the 
holes can be strongly affected by nonlinearities, while the radiation 
reaction due to gravitational wave emission is weak. It is useful to 
approximate such a system as having helical symmetry, in which the 
sources and fields rotate rigidly. Two aspects of this physical problem 
are then important to the mathematical methods explored in this paper: 
(i)~``Helical reduction," that is, imposing helical symmetry, converts 
the PDEs describing the fields into a system of helically reduced wave 
equations (HRWEs), a mixed system that is elliptic inside, and 
hyperbolic outside a type changing surface. (ii)~Nonlinearities are 
significant only interior to this type changing surface. In this paper 
we have divided the topologically nontrivial domain of the problem into 
basic subdomains. Crucially, only a single subdomain contains the type
changing surface and the hyperbolic region exterior to it, the outer
annulus/shell subdomain (annulus for 2d, shell for 3d). The other 
subdomain or subdomains span an elliptic problem. We have therefore 
confined aspect (i) to a topologically simple outer subdomain, and 
aspect (ii) to the subdomains spanning an elliptic problem.

Sections \ref{sec:outerAnnBC} and \ref{sec:sparse} of this paper have 
focused on the model problem of the 
2d HRWE for a linear scalar field, though some generalization to 3d
and to nonlinear problems has been included. Likewise, while the 
numerical results in Sec.~4 have predominately focused on the 2d 
linear scalar problem, we have also included some results for a 
nonlinear 2d scalar model and for a linear 3d scalar model. To obtain 
our numerical results we have necessarily made many choices in the 
details. This was also case in Sec.~3; clarity of presentation required 
some specificity in the adopted methods. It is, however, important to 
distinguish the full set of choices made from what we believe are 
robust conclusions that transcend many of the choices. In the present 
section, we therefore remark on what, in our view, are the central 
lessons of this work.

The choice to use a spectral (or a pseudospectral) method does not 
need justification for an elliptic problem. Our mixed problem 
inherits much of that justification. In particular, we are solving a 
boundary value problem, so a large set of equations must be solved 
simultaneously.  For such problems spectral methods, which reduce 
the number of unknowns, are particularly valuable and widely used. 
A less general issue is the nature of the decomposition of the two 
center domain of the problem into subdomains. The choice illustrated 
in Fig.~\ref{domaindecomp} is very much driven by the convenience of 
rectangular and annular subdomains, for on such regions spectral 
methods are well developed. For analogous reasons, the similar 
decomposition in Fig.~\ref{domaindecomp3} would be used for the 
3d problem. 

What we wish to emphasize here is that one should make a distinction 
between the choices for the outer annulus/shell and those for the 
inner region, the subdomain or concatenation of subdomains for the 
elliptical region. (One could also divide the outer shell/annulus 
into further concentric sub-shells/sub-annuli, but this possibility 
only adds further technical complexity). It is quite conceivable that 
the optimal choice for the outer region is not the same as that for 
the inner region.  One might, for example, treat the inner elliptic 
problem by whatever method most conveniently or efficiently generates 
a set of equations for the interior, and then use the spectral method 
of this paper only for the outer region with its mixed PDEs. Even 
more freedom is available if the inner and outer regions are solved 
separately, with an iteration between them, as was done in the 
second numerical experiment of Sec.~\ref{3dsim}. In this case, 
one could use separate iterations to solve within each region, 
each within the outer loop iteration that feeds values from the
inner to the outer region and {\em vice versa}. We will return to 
this possibility in a moment.

We have made the choice of using integration preconditioning (IPC) 
in all subdomains. Whether or not this is an optimal choice remains
unresolved. IPC is known to be efficient in reducing the
condition number of the coefficient matrix for certain types of ODE 
boundary value problems. Our study of a novel boundary value problem, 
one heretofore not considered in the context of spectral methods, 
has to some extent further elucidated the utility of IPC in such 
ODE settings. The results in Table~\ref{up5150} 
demonstrate that IPC reduces the condition number of the matrix 
stemming from a problem not considered in our primary 
reference\cite{CHHT}. Moreover, our work demonstrates that IPC 
affords a relatively direct way to handle a multidomain scenario in 
two dimensions. (See also \cite{WKC} for an application of IPC in a 
1d multidomain setting.) However, it remains to be seen if its 
advantages will carry over to 3d and to PDEs significantly 
different from those of our test problems.

Coming to grips with whether or not IPC is an appropriate method 
for 3d PSW problems, requires that we ask some broader background 
questions. In particular, is it important to reduce the condition 
number of the coefficient matrix? In Sec.~\ref{numerical}, solutions 
to 2d problems were found by straightforward Gaussian elimination 
for which conditioning issues are not as critical. For 3d problems, 
however, we almost certainly expect to use an iterative Krylov 
method, and GMRES in particular. The efficiency of such iterative 
solvers is known to depend sensitively on condition number. But such 
methods are also sensitive to other aspects of the coefficient 
matrix, in particular to the distribution of eigenvalues\cite{Kelley}. 
In any case, owing to the relevant domain decomposition 
(Fig.~\ref{domaindecomp3})
for a 3d problem, insofar as the elliptic problem is concerned the 
relevant issue is the effectiveness of IPC for the 3d HRWE on 
cylindrical and (inner) spherical--shell domains. As noted in 
\cite{CHHT,Boyd}, special alternate techniques are often necessary 
for polar and cylindrical subdomains which contain a center or 
central axis.

We turn now from the open question of whether IPC could be 
effectively applied to all, or almost all, subdomains of a 3d PSW 
problem, to the rather clearer question of IPC for the outer region 
itself. For a scheme in which inner and outer regions are separately 
solved, we have some enthusiasm for the advantages of IPC. Our 
enthusiasm stems from the fact that in the outer region the Fourier, 
or spherical harmonic modes do not mix for a linear problem. This 
nonmixing originates in the symmetry of the problem, the fact that 
the annulus or shell is concentric with the coordinate surfaces of 
constant radius. The situation is closely analogous to the reason that 
separation of variables applied to this region leads to a set of 
decoupled radial ODEs. In line with theoretical and numerical 
arguments presented in 
\cite{CHHT}, our studies indicate that IPC is a promising way of 
solving the system of ODEs arising in the outer region of our problem.

These simple considerations do not apply, of course, if the problem is
nonlinear.  The nonlinearity will mix Fourier or spherical harmonic
modes. However, the nonlinearities in the underlying physical problem
are very weak in the outer region. (Here we are assuming that the
inner surface of the outer region is chosen to be near the maximum
radius that allows it to be inside the type changing surface and to
have the needed overlap with the subdomains of the interior region.)
As a result, there would seem to be number of possibilities for
handling the nonlinearity. First, it is quite plausible that it is
adequate to treat the problem as linear in the outer region. If it is
not adequate to treat the nonlinearity as null, then one could exploit
the fact that it is weak. The weakly nonlinear problem could be solved
by iteration in which each iteration generates a ``known'' source
term, and that source term is projected onto the angular mode basis
before the system is solved. Just as would be the case with separation
of variables, this procedure would not mix modes (after the projection
of the known source is done) so the solution would again be highly
efficient.  The weakness of the source should guarantee that a few
iterations would be sufficient to get the needed resolution. Finally,
a Newton--Krylov method might be based upon use of the outer domain's
spectral transform and inverse transform (with all nonlinear
operations performed in point space). For a weak nonlinearity the
associated Jacobian should not significantly perturb the basic linear
operator.

Our next steps with these methods will parallel those of
Refs.~\cite{WheKrivPri}--\cite{BBHP2006}.  The method used in those
studies is, by its nature, incapable of high precision, whereas our
approach is capable of the precision that is needed for some purposes
(initial data for evolution codes and studies of radiation reaction
for instance). In particular, we will next formulate 3d linear
and nonlinear scalar model HRWE problems.  We will then
use the techniques of Ref.~\cite{BBP2006} to do linearized gravity,
and those of Ref.~\cite{BBHP2006} to do gravity with a post-Minkowski
approximation. Finally, we will treat the problem in fully nonlinear
general relativity.

\section{Acknowledgments}
For helpful conversations and clarifying remarks, we thank 
E.~A.~Coutsias, T. Hagstrom, J.~S.~Hesthaven, and 
particularly G.~von Winckel. This research was supported 
by NSF grants PHY 0514282 and PHY 0554367 (to UTB), 
ARO DAAD19-03-1-0146 (to UNM), and DMS 0554377 and 
DARPA/AFOSR FA9550-05-1-0108 (to Brown University).

\end{document}